\documentclass[twocolumn,3p,times,procedia]{elsarticle}

\usepackage{ecrc}
\usepackage{graphicx}
\usepackage{textcomp}
\usepackage{xcolor}
\usepackage[utf8]{inputenc}
\usepackage{booktabs} % For form
\usepackage{amsmath}
\usepackage{amsfonts}
\usepackage{amssymb}
\usepackage{amsthm}
\usepackage{algorithm}
\usepackage{algpseudocode}
\renewcommand{\arraystretch}{1}
\usepackage{csquotes}
\usepackage{enumerate}
\usepackage{multirow}
\usepackage{multicol}
\usepackage{xurl}
\usepackage{tabularx}
\usepackage{array}
\newtheorem{definition}{Definition}
\usepackage{xcolor}
\usepackage{subfig}
\usepackage{flushend}
\usepackage{comment}
\usepackage{mathrsfs}
\usepackage{gensymb}
\usepackage{tikz}
\usetikzlibrary{trees}
\usepackage{breqn}
\newtheorem{theorem}{Theorem}
\usepackage{titlesec}
\usepackage{url}
\volume{00}

\firstpage{1}

\runauth{H.H. Arcolezi et al.}

\jid{procs}

\usepackage{caption}

\usepackage{fancyhdr}

\fancyheadoffset[RO,EL]{0pt}
\usepackage{geometry}

\begin{document}
\sloppy

\begin{frontmatter}




\dochead{}
%% e.g. \dochead{17th International Conference on Dynamical Processes in Excited States of Solids}
\title{
\begin{flushleft}
Improving the utility of locally differentially private protocols for longitudinal and multidimensional frequency estimates\tnoteref{label1}
\end{flushleft}
}

\author[]{ \leftline {H\'eber H. Arcolezi$^*$$^{a,b}$, Jean-François Couchot$^b$, Bechara Al Bouna$^c$, Xiaokui Xiao$^d$}}

\address{  \leftline {$^a$Inria and École Polytechnique (IPP), Palaiseau, France}

 \leftline {$^b$Femto-ST Institute, Univ. Bourg. Franche-Comt\'e, UBFC, CNRS, Belfort, France}

  \leftline {$^c$TICKET Lab., Antonine University Hadat-Baabda, Baabda, Lebanon}

 \leftline {$^d$School of Computing, National University of Singapore, Singapore, Singapore}
}
\tnotetext[label1]{Final version accepted in the journal Digital Communications and Networks.
Version of Record: \url{https://doi.org/10.1016/j.dcan.2022.07.003}.}

\cortext[]{Corresponding author (email: heber.hwang-arcolezi@inria.fr).}

\begin{abstract}

\noindent This paper investigates the problem of collecting multidimensional data throughout time (i.e., longitudinal studies) for the fundamental task of frequency estimation under Local Differential Privacy (LDP) guarantees. Contrary to frequency estimation of a single attribute, the multidimensional aspect demands particular attention to the privacy budget. Besides, when collecting user statistics longitudinally, privacy progressively degrades. Indeed, the ``multiple" settings in combination (i.e., many attributes and several collections throughout time) impose several challenges, for which this paper proposes the first solution for frequency estimates under LDP. To tackle these issues, we extend the analysis of three state-of-the-art LDP protocols (Generalized Randomized Response -- GRR, Optimized Unary Encoding -- OUE, and Symmetric Unary Encoding -- SUE) for both longitudinal and multidimensional data collections. While the known literature uses OUE and SUE for two rounds of sanitization (a.k.a. memoization), i.e., L-OUE and L-SUE, respectively, we analytically and experimentally show that starting with OUE and then with SUE provides higher data utility (i.e., L-OSUE). Also, for attributes with small domain sizes, we propose Longitudinal GRR (L-GRR), which provides higher utility than the other protocols based on unary encoding. Last, we also propose a new solution named \underline{A}daptive \underline{L}DP for \underline{LO}ngitudinal and \underline{M}ultidimensional \underline{FRE}quency \underline{E}stimates (ALLOMFREE), which randomly samples a single attribute to be sent with the whole privacy budget and adaptively selects the optimal protocol, i.e., either L-GRR or L-OSUE. As shown in the results, ALLOMFREE consistently and considerably outperforms the state-of-the-art L-SUE and L-OUE protocols in the quality of the frequency estimates.

\end{abstract}


\end{frontmatter}

\section{Introduction}\label{sec:intro}

\subsection{Background}

In recent years, Differential Privacy (DP)~\cite{Dwork2006,Dwork2006DP} has been increasingly accepted as the current standard for data privacy~\cite{dwork2014algorithmic,aktay2020google,linkedin,DL_DP}. In the centralized model of DP, a trusted curator has access to the entire raw data of users (e.g., the Census Bureau~\cite{census,census2021}). By ``trusted", we mean that curators do not misuse or leak private information of individuals. However, this assumption does not always hold in real life, e.g., data breaches are all too common~\cite{data_breaches}. 

To preserve privacy at the user-side, an alternative approach, namely, Local Differential Privacy (LDP), was initially formalized in~\cite{first_ldp}. With LDP, rather than trust a data curator to have the raw data and sanitize it to output queries, each user applies a DP mechanism to their data before transmitting it to the data collector server. The local DP model allows collecting data in unprecedented ways and, therefore, it has been widely adopted by industry (e.g., Google Chrome browser~\cite{rappor}, Microsoft windows 10 operation system~\cite{microsoft}, Apple iOS and macOS~\cite{apple}).

\subsection{Motivation and problem statement} \label{sub:problem_statement}

When collecting data in practice, one is often interested in multiple attributes of a population, i.e., \textit{multidimensional data}. For instance, in crowd-sourcing applications, the server may collect both demographic information (e.g., gender, nationality) and user habits in order to develop personalized solutions for specific groups. In addition, one generally aims to collect data from the same users throughout time (i.e., \textit{longitudinal} studies), which is essential in many situations~\cite{rappor,microsoft}. For example, the fact that two medical acts identified at a different time have been performed on the same patient, or two different patients mean treatment in the first case or two isolated acts in the second. 

So, in this paper, we focus on the problem of private frequency (or histogram) estimation of multiple attributes throughout time with LDP. Frequency estimation is a primary objective of LDP, in which the data collector (a.k.a. the aggregator) decodes all the privatized data of the users and then estimates the number of users for each possible value. More formally, we assume there are $d$ attributes $A=\{A_1,A_2,...,A_d\}$, where each attribute $A_j$ with a discrete domain has a specific number of value $k_j=|A_j|$. Each user $u_i$ for $i \in \{1,2,...,n\}$ has a tuple $\textbf{v}^{(i)}=(v^{(i)}_{1},v^{(i)}_{2},...,v^{(i)}_{d})$, where $v^{(i)}_{j}$ represents the value of attribute $A_j$ in record $\textbf{v}^{(i)}$. Thus, for each attribute $A_j$ at time $t \in [1,\tau]$, the aggregator's goal is to estimate a $k_j$-bins histogram, including the frequency of all values in $A_j$. 

Indeed, in both longitudinal and multidimensional settings, one needs to consider the allocation of the privacy budget, which can grow extremely quickly due to the composition theorem~\cite{dwork2014algorithmic}. However, on the one hand, most academic literature on frequency estimation~\cite{tianhao2017,Cormode2021,Murakami2019,Wang2017,kairouz2016discrete,Hadamard,Alvim2018,Zhao2019,Li2020} focuses on a single data collection (i.e., non-longitudinal studies). On the other hand, the studies for collecting multidimensional data with LDP mainly focus on other complex tasks (e.g., analytical/range queries~\cite{Xu2020,Jianyu2020,Gu2019,Cormode2019}, estimating marginals~\cite{Shen2021,Peng2019,Zhang2018,Ren2018,Fanti2016}) or numerical data only (e.g.,~\cite{xiao2,wang2019,Duchi2018,Wang2021_b}). 

\subsection{Summary of contributions} \label{sub:contributions}

In this paper, we extend the analysis of three state-of-the-art LDP protocols, namely, Generalized Randomized Response (GRR)~\cite{kairouz2016discrete}, Optimized Unary Encoding (OUE)~\cite{tianhao2017}, and Symmetric Unary Encoding (SUE)~\cite{rappor} for both longitudinal and multidimensional frequency estimates. On the one hand, for all three protocols, we theoretically prove that randomly sampling a single attribute per user improves data utility, which is an extension of common results in the LDP literature~\cite{erlingsson2020encode,Jianyu2020,Wang2021,Zhang2018,Arcolezi2021}. 

On the other hand, in the literature, both SUE and OUE protocols have been extended (and also applied~\cite{Kim2018,Vidal2020}) to longitudinal studies based on the concept of \textit{memoization}~\cite{rappor,microsoft}, i.e., L-SUE and L-OUE, respectively. However, we numerically and experimentally show that combining both protocols provides higher data utility, i.e., starting with OUE and then with SUE (L-OSUE) optimizes data utility better than using SUE or OUE twice. In addition, we also extend GRR for longitudinal studies (i.e., L-GRR), which provides higher data utility than the other protocols based on unary encoding for attributes with a small domain size. 

Lastly, in a multidimensional setting having different domain sizes for each attribute, a dynamic selection of longitudinal LDP protocols is preferred. Therefore, we propose a new solution named \underline{A}daptive \underline{L}DP for \underline{LO}ngitudinal and \underline{M}ultidimensional \underline{FRE}quency \underline{E}stimates (ALLOMFREE), which combines all the aforementioned results. More specifically, ALLOMFREE randomly samples a single attribute to be sent with the whole privacy budget and adaptively selects the optimal protocol, i.e., either L-GRR or L-OSUE. To validate our proposal, we conduct a comprehensive and extensive set of experiments on four real-world open datasets. Under the same privacy guarantee, results show that ALLOMFREE consistently and considerably outperforms the state-of-the-art L-SUE and L-OUE protocols in the quality of the frequency estimates.

The remainder of this paper is organized as follows. In Section~\ref{sec:background}, we review the privacy notion in consideration, i.e., LDP and the protocols. In Section~\ref{sec:multidimensional}, we extend the analysis of GRR, OUE, and SUE to multidimensional data collections. In Section~\ref{sec:longitudinal} we present the \textit{memoization}-based framework for longitudinal data collections, the extension and analysis of longitudinal GRR and the longitudinal UE-based protocols and the numerical evaluation of their performance, and we present our ALLOMFREE solution. In Section~\ref{sec:results_discussion}, we present experimental results and discuss our results. In Section~\ref{sec:rel_work} we review the related work. Lastly, in Section~\ref{sec:conc}, we present the concluding remarks and future directions.

\section{Theoretical background}\label{sec:background}

In this section, we briefly present the concept of privacy considered in this work, that is, LDP, and the LDP protocols we will apply in this paper.

\subsection{LDP}\label{sub:ldp}

Local differential privacy, initially formalized in~\cite{first_ldp}, protects an individual's privacy during the data collection process. A formal definition of LDP is given as follows:

\begin{definition}[$\epsilon$-Local Differential Privacy]\label{def:ldp} A randomized algorithm ${\mathcal{A}}$ satisfies $\epsilon$-LDP if, for any pair of input values $v_1, v_2 \in Domain(\mathcal{A})$ and any possible output $y$ of ${\mathcal{A}}$:

\begin{equation*}
    \Pr[{\mathcal{A}}(v_1) = y]\leq e^{\epsilon }\cdot \Pr[{\mathcal{A}}(v_2) = y] 
\label{eq:ldp}
\end{equation*}

\end{definition}

Similar to the centralized model of DP, LDP also enjoys several important properties, e.g., immunity to post-processing ($F(\mathcal{A})$ is $\epsilon$-LDP for any function $F$) and composability~\cite{dwork2014algorithmic}. That is, combining the results from $d$ locally differentially private protocols also satisfy LDP. If these protocols are applied separately in disjointed subsets of the dataset, $\epsilon = max(\epsilon_1$-, \ldots,  $\epsilon_d)$-LDP (parallel composition). On the other hand, if these protocols are sequentially applied to the same dataset, $\epsilon = \sum_{i=1}^{d}\epsilon_i$-LDP (sequential composition).

\subsection{LDP protocols}\label{sub:ld_mechanisms}

Randomized Response (RR), a surveying technique proposed by Warner~\cite{Warner1965}, has been the building block for many LDP protocols. Let $A_j=\{v_1,v_2,...,v_{k_j}\}$ be a set of $k_j=|A_j|$ values of a given attribute and let $\epsilon$ be the privacy budget, we review three state-of-the-art LDP mechanisms for single-frequency estimation (a.k.a. frequency oracles) that will be used in this paper.

\subsubsection{GRR} \label{subsub:grr}

The \textit{k}-Ary RR~\cite{kairouz2016discrete} mechanism extends RR to the case of $k_j \geq 2$ and is also referred to as direct encoding~\cite{tianhao2017} or Generalized RR (GRR)~\cite{Wang2018,Wang2020_post_process,Zhang2018}. Throughout this paper, we use the term GRR for this LDP protocol. Given a value $v \in A_j$, \textit{GRR($v$)} outputs the true value with probability $p$, and any other value $v'\in A_j$ such that $v' \neq v$ with probability $1-p$. More formally, the perturbation function is defined as:

\begin{equation*}
    \forall{y \in A_j} \Pr [\mathcal{A}_{GRR(\epsilon)}(v)=y] = \begin{cases} p=\frac{e^{\epsilon}}{e^{\epsilon}+k_j-1} , \textrm{ if } y = v\\ q=\frac{1}{e^{\epsilon}+k_j-1}, \textrm{ if } y \neq v  \end{cases}
\end{equation*}

This satisfies $\epsilon$-LDP since $\frac{p}{q}=e^{\epsilon}$. On expectation, the number of times that a value $v_i$ is reported, $N_i$, for $i \in [1,k_j]$, is given by:
    
\begin{equation*}
    \mathbb{E}[N_i] = n f(v_i) p + n (1 - f(v_i))q 
\end{equation*}

\noindent in which $N_i$ is the number of times the value $v_i$ has been reported, $f(v_i)$ is the real frequency of value $v_i$, and $n$ is the total number of users. This immediately provides the normalized estimation $\hat{f}(v_i)$ that each value $v_i$ occurs as~\cite{kairouz2016discrete,tianhao2017,rappor}:

\begin{equation}\label{eq:est_pure}
    \hat{f}(v_i) = \frac{N_i - nq}{n(p - q)}
\end{equation}

In~\cite{tianhao2017}, the authors prove that $\hat{f}(v_i)$ in Eq.~\eqref{eq:est_pure} is an unbiased estimation of the true frequency $f(v_i)$, and the variance of this estimation is $Var[\hat{f}(v_i)]= \frac{q(1-q)}{n(p-q)^2} + \frac{f(v_i)(1-p-q)}{n(p-q)}$. In the case of small $f(v_i) \sim 0$, this variance is dominated by the first term, which gives the \textit{approximate} variance as~\cite{tianhao2017}:

\begin{equation}\label{eq:var_pure}
    Var^*[\hat{f}(v_i)] = \frac{q (1 - q)}{n(p - q)^2} 
\end{equation}

Since the estimation in Eq.~\eqref{eq:est_pure} is unbiased, its variance $Var[\hat{f}(v_i)]$ is equal to the Mean Squared Error (MSE), which is commonly used as an accuracy metric (e.g., cf.~\cite{Wang2020_post_process,Wang2021_b}) and also adopted in this paper. Replacing $p=\frac{e^{\epsilon}}{e^{\epsilon}+k_j-1}$ and $q=\frac{1}{e^{\epsilon}+k_j-1}$ into Eq.~\eqref{eq:var_pure}, the GRR variance is calculated as:

\begin{equation}\label{eq:var_grr}
    Var^*[\hat{f}_{GRR}(v_i)] = \frac{e^{\epsilon} + k_j - 2}{n(e^{\epsilon}-1)^2}
\end{equation}

\subsubsection{Unary encoding-based}\label{subsub:UE}

Protocols based on Unary Encoding (UE) consist of transforming a value $v$ into a binary representation of it. So, first, for a given value $v$, $B=UE(v)$, where $B=[0,0,...,1,0,...0]$, a $k_j$-bit array where only the $v$-th position is set to one. Next, the bits $i$, for $i \in [1,k_j]$, from $B$ are flipped, depending on parameters $p$ and $q$, to generate a sanitized vector $B'$, in which:

\begin{equation*}
    \Pr[B'_i=1] =\begin{cases} p, \textrm{ if } B_i=1 \\ q, \textrm{ if } B_i=0 \end{cases}
\end{equation*}

The proof that the UE-based protocols satisfy $\epsilon$-LDP for

\begin{equation}\label{eq:eps_UE}
    \epsilon = ln\left( \frac{p(1-q)}{(1-p)q} \right )
\end{equation}

\noindent is known in the literature and can be found in~\cite{rappor,tianhao2017}. In~\cite{tianhao2017} the authors presented two ways for selecting probabilities $p$ and $q$, which determines the protocol variance. One well-known UE-based protocol is the basic one-time RAPPOR~\cite{rappor}, referred to as Symmetric UE (SUE), which selects $p=\frac{e^{\epsilon/2}}{e^{\epsilon/2}+1}$ and $q=\frac{1}{e^{\epsilon/2}+1}$, where $p+q=1$ (symmetric). The estimated frequency $\hat{f}(v_i)$ that a value $v_i$ occurs for $i \in [1,k_j]$ is also calculated using Eq.~\eqref{eq:est_pure}. Replacing $p=\frac{e^{\epsilon/2}}{e^{\epsilon/2}+1}$ and $q=\frac{1}{e^{\epsilon/2}+1}$ into Eq.~\eqref{eq:var_pure}, the SUE variance is calculated as~\cite{rappor}:

\begin{equation}\label{eq:var_sue}
    Var^*[\hat{f}_{SUE}(v_i)] = \frac{e^{\epsilon/2}}{n(e^{\epsilon/2}-1)^2}
\end{equation}

Moreover, rather than select $p$ and $q$ to be symmetric, Wang \textit{et al}.~\cite{tianhao2017} proposed Optimized UE (OUE), which selects parameters $p=\frac{1}{2}$ and $q=\frac{1}{e^{\epsilon}+1}$ that minimize the variance of UE-based protocols while still satisfying $\epsilon$-LDP. Similarly, the estimation method used in Eq.~\eqref{eq:est_pure} equally applies to OUE. Replacing $p=\frac{1}{2}$ and $q=\frac{1}{e^{\epsilon}+1}$ into Eq.~\eqref{eq:var_pure}, the OUE variance is calculated as~\cite{tianhao2017}:

\begin{equation}\label{eq:var_oue}
   Var^*[\hat{f}_{OUE}(v_i)] = \frac{4e^{\epsilon}}{n(e^{\epsilon}-1)^2} 
\end{equation}

\section{Multidimensional frequency estimates with LDP}\label{sec:multidimensional}

In the literature, few work for collecting multidimensional data with LDP is based on random sampling (i.e., dividing users in groups)~\cite{xiao2,wang2019,Duchi2018,Wang2021_b,tianhao2017,Arcolezi2021}. This technique reduces both dimensionality and communication costs, which will also be the focus of this paper. Let $d\geq2$ be the total number of attributes, $\textbf{k}=[k_1,k_2,...,k_d]$ be the domain size of each attribute, $n$ be the number of users, and $\epsilon$ be the privacy budget. An intuitive solution (\textit{Spl}) is to split the privacy budget, i.e., assigning $\epsilon/d$ for each attribute. The other solution (\textit{Smp}) is based on uniformly sampling (without replacement) only $r$ attribute(s) out of $d$ possible ones, i.e., assigning $\epsilon/r$ per attribute. Notice that both solutions satisfy $\epsilon$-LDP according to the sequential composition theorem~\cite{dwork2014algorithmic}.

For the first case, \textit{Spl}, the variances ($\sigma^{2}_{1}$) of GRR, SUE, and OUE are respectively:

\begin{equation} \label{eq:var_spl}
\begin{split}
\sigma^{2}_{1,GRR} & = \frac{e^{\epsilon/d} + k_j - 2}{n(e^{\epsilon/d}-1)^2} \\
\sigma^{2}_{1,SUE} & = \frac{e^{\epsilon/2d}}{n(e^{\epsilon/2d}-1)^2}    \\
\sigma^{2}_{1,OUE} & = \frac{4e^{\epsilon/d}}{n(e^{\epsilon/d}-1)^2} 
\end{split}
\end{equation}

For the second case, \textit{Smp}, the number of users per attribute is reduced to $nr/d$. Thus, the variances ($\sigma^2_2$) of GRR, SUE, and OUE are, respectively:

\begin{equation} \label{eq:var_smp}
\begin{split}
\sigma^{2}_{2,GRR} & = \frac{d(e^{\epsilon/r} + k_j - 2)}{nr(e^{\epsilon/r}-1)^2} \\
\sigma^{2}_{2,SUE} & = \frac{d(e^{\epsilon/2r})}{nr(e^{\epsilon/2r}-1)^2} \\
\sigma^{2}_{2,OUE} & = \frac{d(4e^{\epsilon/r})}{nr(e^{\epsilon/r}-1)^2} 
\end{split}
\end{equation}

Notice that if $r=d$ in Eq.~\eqref{eq:var_smp}, one achieves Eq.~\eqref{eq:var_spl}. Practically, the objective is reduced to finding $r$, which minimizes $\sigma^2_2$ for each protocol. In this way, to find the optimal $r$ for each protocol, we first multiply each $\sigma^2_2$ in Eq.~\eqref{eq:var_smp} by $\epsilon$. Without losing generality, minimizing $\sigma^{2}_{2,GRR}$, $\sigma^{2}_{2,SUE}$, and $\sigma^{2}_{2,OUE}$ is equivalent to minimizing $\frac{\epsilon e^{\epsilon/r}}{r(e^{\epsilon/r}-1)^2}$, $\frac{\epsilon e^{\epsilon/2r}}{r(e^{\epsilon/2r}-1)^2}$, and $\frac{\epsilon e^{\epsilon/r}}{r(e^{\epsilon/r}-1)^2}$, respectively. Hence, let $x=r/\epsilon$ be the independent variable, 
$\sigma^2_{2,GRR}$ and 
$\sigma^2_{2,OUE}$ can be rewritten as $y_1=\frac{1}{x}\cdot \frac{e^{1/x}}{(e^{1/x}-1)^2}$, and $\sigma^2_{2,SUE}$ can be rewritten as $y_2=\frac{1}{x}\cdot \frac{e^{1/2x}}{(e^{1/2x}-1)^2}$ as functions over $x$. It is not hard to prove that both $y_1$ and $y_2$ are increasing functions w.r.t. $x$. Therefore, the minimum and optimal number of attributes per user is $r=1$ for all three protocols. We highlight that this is a common result in the LDP literature obtained for different protocols and contexts~\cite{xiao2,wang2019,Wang2021_b,tianhao2017,Jianyu2020,Wang2021,erlingsson2020encode,bassily2017practical}. 

\textbf{Therefore, in this paper, we adopt the multidimensional setting \textit{Smp} with $r=1$}. In this setting, users tell the data collector whose attribute is sampled, and its perturbed value ensures $\epsilon$-LDP by applying either GRR or UE-based protocols; the data analyst server would not receive any information about the remaining $d-1$ attributes. 

\section{Longitudinal frequency estimates with LDP}\label{sec:longitudinal}

In this section, we first present the \textit{memoization}-based framework for longitudinal data collections. Next, we present the analysis of longitudinal GRR and longitudinal UE-based protocols. Lastly, we numerically evaluate the extended longitudinal protocols and propose our ALLOMFREE solution.

\subsection{Memoization-based data collection with LDP} \label{sub:memoization}

In the literature, many studies focus on how to collect and analyze categorical data longitudinally based on \textit{memoization}~\cite{rappor,microsoft,erlingsson2020encode}. The key idea behind memoization is using two sanitization processes. The first round ($RR_1$) replaces the real value $B$ with a sanitized one $B'$ with a higher epsilon ($\epsilon_{\infty}$). Whenever one intends to report $B$, $B'$ shall be reused to produce other sanitized versions $B''$ with lower epsilon values. Notice that the second sanitization ($RR_2$) is a \textit{must} to avoid ``averaging attacks", in which adversaries can reconstruct the true value from multiple sanitized versions of it. This technique allows achieving privacy over time with an upper bound value of $\epsilon_{\infty}$-LDP.

Let $A_j=\{v_1,v_2,...,v_{k_j}\}$ be a set of $k_j=|A_j|$ values of a given attribute and let $\epsilon$ be the privacy budget. In this paper, for both $RR_1$ and $RR_2$ steps, we will apply either GRR, SUE, or OUE. The unbiased estimator in Eq.~\eqref{eq:est_pure} for the frequency $f(v_i)$ of each value $v_i$ for $i \in [1,k_j]$ is now extended to: 

\begin{equation}\label{eq:est_longitudinal}
    \hat{f}_L(v_i) = \frac{\frac{N_i - nq_2}{(p_2-q_2)} - nq_1}{n(p_1-q_1)} = \frac{N_i - nq_1(p_2-q_2) - nq_2}{n(p_1-q_1)(p_2-q_2)} 
\end{equation}

\noindent in which $N_i$ is the number of times the value $v_i$ has been reported, $n$ is the total number of users, $p_1$ and $q_1$ are the parameters used by an LDP protocol for $RR_1$, and $p_2$ and $q_2$ are the parameters used by an LDP protocol for $RR_2$. Eq.~\eqref{eq:est_longitudinal} is the result of using the unbiased estimator of Eq.~\eqref{eq:est_pure} with two rounds of sanitization.

\begin{theorem} \label{theo:est_long} The estimation result $\hat{f}_L(v_i)$ in Eq.~\eqref{eq:est_longitudinal} is an unbiased estimation of $f (v_i)$ for any value $v_i \in A_j$.
\end{theorem}

\noindent \textit{Proof.}

\begin{equation*}
\begin{aligned}
    \mathbb{E}[\hat{f}_L(v_i)] &= \mathbb{E}\left[ \frac{N_i - nq_1(p_2-q_2) - nq_2}{n(p_1-q_1)(p_2-q_2)} \right] \\
    &= \frac{\mathbb{E}[N_i]- nq_1(p_2-q_2) - nq_2}{n(p_1-q_1)(p_2-q_2)}  
\end{aligned}
\end{equation*}

Let us focus on 

\begin{equation*}
\begin{aligned}
    \mathbb{E}[N_i] &= n f(v_i) \left(p_{1} p_{2} + q_{2} \left(1 - p_{1}\right)\right) \\
    &+ n \left(1 - f(v_i)\right) \left(p_{2} q_{1} + q_{2} \left(1 - q_{1}\right)\right)
\end{aligned}
\end{equation*}

Thus,

\begin{equation*}
\begin{aligned}
    \mathbb{E}[\hat{f}_L(v_i)] &= \\
    & \frac{n f(v_i) \left(p_{1} p_{2} + q_{2} \left(1 - p_{1}\right)\right) - n q_{1} \left(p_{2} - q_{2}\right) - n q_{2}}{n \left(p_{1} - q_{1}\right) \left(p_{2} - q_{2}\right)}\\
    &+ \frac{\left(- f(v_i) n + n\right) \left(p_{2} q_{1} + q_{2} \left(1 - q_{1}\right)\right)}{n \left(p_{1} - q_{1}\right) \left(p_{2} - q_{2}\right)}  \\
    &= f(v_i)
\end{aligned}
\end{equation*}

\begin{theorem} \label{theo:variance_grr} The variance of the estimation in Eq.~\eqref{eq:est_longitudinal} is:

\begin{equation}\label{var:longitudinal}
\begin{gathered}
    Var[\hat{f}_L(v_i)]  = \frac{\gamma (1-\gamma)}{n (p_1-q_1)^2 (p_2-q_2)^2} \textrm{, where} \\
    \gamma = f(v_i) \left( 2 p_{1} p_{2} - 2 p_{1} q_{2} + 2 q_{2} - 1 \right) + p_{2} q_{1} + q_{2} (1 - q_{1}) 
\end{gathered}
\end{equation}

\end{theorem}

\noindent \textit{Proof.} Thanks to Eq.~\eqref{eq:est_longitudinal}, we have

\begin{equation*}
Var\left[\hat{f}_L(v_i)\right] = 
\frac{Var[N_i]}{n^2 (p_1-q_1)^2 (p_2-q_2)^2}  
\end{equation*}

Since $N_i$ is the number of times the value $v_i$ is observed, it can be defined as $N_i = \sum_{z=1}^n X_z$, where $X_z$ is equal to 1 if the user $z$, 
$1 \le z \le n$ reports value $v_i$, and 0 otherwise. We thus have 
$
Var[N_i] 
= \sum_{z=1}^n Var[X_z] 
= n Var[X]$. Since all the users are independent,

\begin{equation*}
\begin{gathered}
\Pr[X = 1] = P[X^2 = 1] = f(v_i) \left( 2 p_{1} p_{2} - 2 p_{1} q_{2} + 2 q_{2} - 1 \right) \\
+ p_{2} q_{1} + q_{2} (1 - q_{1}) = \gamma 
\end{gathered}
\end{equation*}

We thus have $Var[X]= \gamma - \gamma^2 = \gamma(1 - \gamma) $ and, finally,

\begin{equation*} \label{var:generic}
Var[\hat{f}_L(v_i)] =
\frac{\gamma (1-\gamma)}{n (p_1-q_1)^2 (p_2-q_2)^2}
\end{equation*}

In this work, we will use the \textit{approximate variance}, in which $f(v_i)=0$ in Eq.~\eqref{var:longitudinal}, which gives:

\begin{equation}\label{var:aprox_longitudinal}
\begin{aligned}
    Var^*\left[\hat{f}_L(v_i)\right]  &= \\ &\frac{\left(p_{2} q_{1} - q_{2} \left(q_{1} - 1\right)\right) \left(- p_{2} q_{1} + q_{2} \left(q_{1} - 1\right) + 1\right)}{n (p_1-q_1)^2 (p_2-q_2)^2} 
\end{aligned}
\end{equation}

\subsection{Longitudinal GRR (L-GRR): definition and $\epsilon$-LDP study}\label{subsub:l_de}

Let $V=\{v_1,v_2,...,v_{k_j}\}$ be a set of $k_j$ values of a given attribute and let $v_i$ be the real value. We now describe an extension of GRR for longitudinal studies; we refer to this protocol as L-GRR for the rest of this paper. First, $Encode(v_i)=v_i$ (direct encoding). Next, there are two rounds of sanitization, $RR_1$ and $RR_2$ applying GRR, as described in the following equations.

\begin{enumerate}
    \item $RR_1[GRR]$: Memoize a value $B'$ such that
    \begin{equation*}
    B'=
    \begin{cases}
      v_i, & \text{with probability}\ p_1 \\
      v_{k\neq v_i}, & \text{with probability}\ q_1=\frac{1-p_1}{k_j-1}  \\
    \end{cases}
  \end{equation*}

  \noindent in which $p_1$ and $q_1$ control the level of longitudinal $\epsilon_{\infty}$-LDP. The value $B'$ shall be reused as the basis for all future reports on the real value $v_i$.
  \item $RR_2[GRR]$: Generate a reporting $B''$ such that
  \begin{equation*}\label{eq:perm}
    B''=
    \begin{cases}
      B', & \text{with probability}\ p_2 \\
      v_{k\neq B'}, & \text{with probability}\ q_2=\frac{1-p_2}{k_j-1}  \\
    \end{cases}
  \end{equation*}
  \noindent in which $B''$ is the report to be sent to the server.
\end{enumerate}

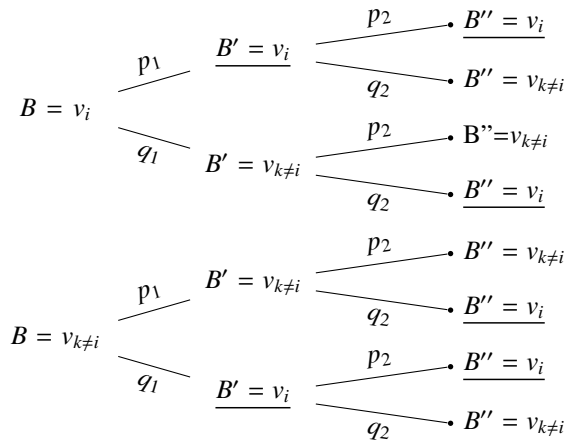
\begin{figure}[t]
\centering
\tikzstyle{level 1}=[level distance=2.6cm, sibling distance=1.5cm]
\tikzstyle{level 2}=[level distance=2.6cm, sibling distance=0.75cm]
\tikzstyle{bag} = [text width=4em, text centered]
\tikzstyle{end} = [circle, minimum width=2pt,fill, inner sep=0pt]
\begin{tikzpicture}[grow=right, sloped]
\node[bag] {$B=v_i$}
child {
node[bag] {$B'=v_{k\neq i}$}
child {
node[end, label=right:
{\underline{$B''=v_i$}}] {}
edge from parent
% node[above] {4}
node[below] {$q_2$}
}
child {
node[end, label=right:
{B''=$v_{k\neq i}$}] {}
edge from parent
node[above] {$p_2$}
% node[below] {8}
}
edge from parent
% node[above] {}
node[below] {$q_1$}
}
child {
node[bag] {\underline{$B'=v_{i}$}}
child {
node[end, label=right:
{$B''=v_{k\neq i}$}] {}
edge from parent
% node[above] {13}
node[below] {$q_2$}
}
child {
node[end, label=right:
{\underline{$B''=v_i$}}] {}
edge from parent
node[above] {$p_2$}
% node[below] {17}
}
edge from parent
node[above] {$p_1$}
% node[below] {19}
};
\end{tikzpicture}
\begin{tikzpicture}[grow=right, sloped]
\node[bag] {$B=v_{k\neq i}$}
child {
node[bag] {\underline{$B'=v_{i}$}}
child {
node[end, label=right:
{$B''=v_{k\neq i}$}] {}
edge from parent
% node[above] {4}
node[below] {$q_2$}
}
child {
node[end, label=right:
{\underline{$B''=v_{i}$}}] {}
edge from parent
node[above] {$p_2$}
% node[below] {8}
}
edge from parent
% node[above] {}
node[below] {$q_1$}
}
child {
node[bag] {$B'=v_{k\neq i}$}
child {
node[end, label=right:
{\underline{$B''=v_{i}$}}] {}
edge from parent
% node[above] {13}
node[below] {$q_2$}
}
child {
node[end, label=right:
{$B''=v_{k\neq i}$}] {}
edge from parent
node[above] {$p_2$}
% node[below] {17}
}
edge from parent
node[above] {$p_1$}
% node[below] {19}
};
\end{tikzpicture}
\caption{Probability trees for two rounds of sanitization using GRR (L-GRR).} \label{fig:tree_l_grr}
\end{figure}

Visually, Fig.~\ref{fig:tree_l_grr} illustrates the probability tree of the L-GRR protocol. In the first round of sanitization, $RR_1$, our proposed L-GRR applies GRR with $p_1=\Pr[ B'=v_i | B=v_i ] =\frac{e^{\epsilon_{\infty}}}{e^{\epsilon_{\infty}}+k_j-1}$ and $q_1=\Pr[ B'=v_i | B=v_{k\neq i} ] =\frac{1-p_1}{k_j-1}=\frac{1}{e^{\epsilon_{\infty}}+k_j-1}$ (\underline{underlined} in the middle of Fig.~\ref{fig:tree_l_grr}), where $k_j=|A_j|$. As discussed in subsection~\ref{subsub:grr}, this \textit{permanent} memoization satisfies $\epsilon_{\infty}$-LDP since $\frac{p_1}{q_1}=e^{\epsilon_{\infty}}$, which is the upper bound. 

On the other hand, with a single collection of data, the attacker's knowledge of $v_i$ comes only from $B''$, which is generated using two randomization steps with GRR. This provides a higher level of privacy protection~\cite{rappor}. From Fig.~\ref{fig:tree_l_grr}, we can obtain the following conditional probabilities:

\begin{equation*}
    \Pr[ B'' | B ] = 
    \begin{cases}
        \Pr[ B''=v_i | B=v_i ] = p_1 p_2 + q_1 q_2 \\
        \Pr[ B''=v_{k\neq i} | B=v_i ] =  p_1 q_2 + q_1 p_2\\
        \Pr[ B''=v_i | B=v_{k\neq i} ] = p_1 q_2 + q_1 p_2 \\
        \Pr[ B''=v_{k\neq i} | B=v_{k\neq i} ] = p_1 p_2 + q_1 q_2 \\
    \end{cases}
\end{equation*}

Let $p_s=\Pr[ B''=v_i | B=v_i ]$ and $q_s = \Pr[ B''=v_i | B=v_{k\neq i} ]$ (\underline{underlined} in the far right of Fig.~\ref{fig:tree_l_grr}), with the second round of sanitization, $RR_2[GRR]$, our proposed L-GRR protocol satisfies $\epsilon_1$-LDP since $\frac{p_s}{q_s}=e^{\epsilon_1}$. Notice that $\epsilon_{1}$ corresponds to a single report (lower bound) and its extension to infinity reports is limited by $\epsilon_{\infty}$ (upper bound) since $RR_2[GRR]$ uses as input the output of $RR_1[GRR]$. More specifically, the calculus of $\epsilon_1$ for L-GRR is:

\begin{equation}\label{eq:e1_grr}
     \epsilon_{1} = \ln{  \left ( \frac{p_1 p_2 + q_1 q_2}{p_1 q_2 + q_1 p_2}\right)}  
\end{equation}

\noindent in which $p_1=\frac{e^{\epsilon_{\infty}}}{e^{\epsilon_{\infty}}+k_j-1}$, $q_1=\frac{1-p_1}{k_j-1}$, and both $p_2$ and $q_2$ are selectable according to $\epsilon_{\infty}$, $\epsilon_{1}$, and $k_j$, calculated as:

\begin{equation} \label{eq:p2_lgrr}
\begin{gathered}
    p_2 = \frac{e^{\epsilon_{1} + \epsilon_{\infty}} - 1}{- k_j e^{\epsilon_{1}} + \left(k_j - 1\right) e^{\epsilon_{\infty}} + e^{\epsilon_{1}} + e^{\epsilon_{1} + \epsilon_{\infty}} - 1}  \\
    q_2  = \frac{1-p_2}{k_j - 1} 
\end{gathered}
\end{equation}

The estimated frequency $\hat{f}_L(v_i)$ that a value $v_i$ occurs for $i \in [1,k_j]$ is calculated using Eq.~\eqref{eq:est_longitudinal}. Lastly, one can calculate the L-GRR approximate variance by replacing the resulting $p_1,q_1,p_2,q_2$ parameters into Eq.~\eqref{var:aprox_longitudinal}.

\subsection{Longitudinal UE (L-UE): definition and $\epsilon$-LDP study}\label{subsub:l_ue}

We now describe the UE-based protocol for longitudinal studies. We refer to this protocol as L-UE for the rest of this paper. Let $V=\{v_1,v_2,...,v_{k_j}\}$ be a set of $k_j$ values of a given attribute and let $v_i$ be the real value. First, $Encode(v_i)=B$ (unary encoding), where $B=[0,0,...,1,0,...0]$, a $k_j$-bit array where only the $v$-th position is set to one. Next, there are two rounds of sanitization, $RR_1$ and $RR_2$, which apply the UE-based protocols, described as follows.

\begin{enumerate}
    \item $RR_1[UE]$: For each bit $i$, $1\le i \le k_j$ in $B$, memoize a value $B'$ such that
    \begin{equation*}
    \Pr[B'_i=1]=
    \begin{cases}
      p_1, & \text{if}\ B_i=1 \\
      q_1, & \text{if}\ B_i=0 
    \end{cases}
  \end{equation*}
  \noindent in which $p_1$ and $q_1$ control the level of longitudinal $\epsilon_{\infty}$-LDP. The value $B'$ shall be reused as the basis for all future reports on the real value $v_i$.
  \item $RR_2[UE]$: For each bit $i$, $1\le i \le k_j$ in $B'$, generate a reporting $B''$ that
  \begin{equation*}
    \Pr[B''_i=1]=
    \begin{cases}
      p_2, & \text{if}\ B'_i=1 \\
      q_2, & \text{if}\ B'_i=0 
    \end{cases}
  \end{equation*}
  \noindent in which $B''$ is the report to be sent to the server.
\end{enumerate}

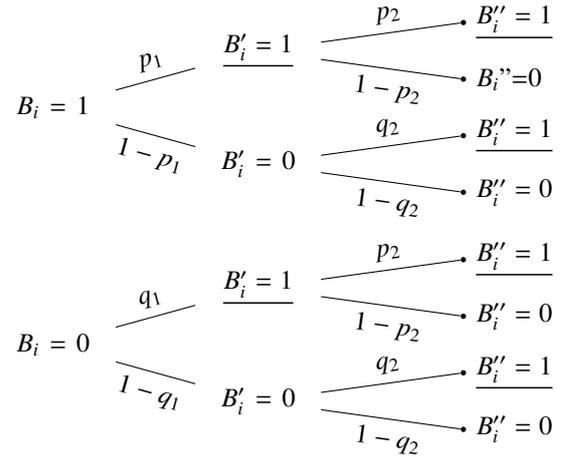
\begin{figure}[t]
\centering
\tikzstyle{level 1}=[level distance=2.7cm, sibling distance=1.5cm]
\tikzstyle{level 2}=[level distance=2.7cm, sibling distance=0.75cm]
\tikzstyle{bag} = [text width=4em, text centered]
\tikzstyle{end} = [circle, minimum width=2pt,fill, inner sep=0pt]
\begin{tikzpicture}[grow=right, sloped]
\node[bag] {$B_i=1$}
child {
node[bag] {$B_i'=0$}
child {
node[end, label=right:
{$B_i''=0$}] {}
edge from parent
% node[above] {4}
node[below] {$1-q_2$}
}
child {
node[end, label=right:
{\underline{$B_i''=1$}}] {}
edge from parent
node[above] {$q_2$}
% node[below] {8}
}
edge from parent
% node[above] {}
node[below] {$1-p_1$}
}
child {
node[bag] {\underline{$B_i'=1$}}
child {
node[end, label=right:
{$B_i$''=0}] {}
edge from parent
% node[above] {13}
node[below] {$1-p_2$}
}
child {
node[end, label=right:
{\underline{$B_i''=1$}}] {}
edge from parent
node[above] {$p_2$}
% node[below] {17}
}
edge from parent
node[above] {$p_1$}
% node[below] {19}
};
\end{tikzpicture}
\begin{tikzpicture}[grow=right, sloped]
\node[bag] {$B_i=0$}
child {
node[bag] {$B_i'=0$}
child {
node[end, label=right:
{$B_i''=0$}] {}
edge from parent
% node[above] {4}
node[below] {$1-q_2$}
}
child {
node[end, label=right:
{\underline{$B_i''=1$}}] {}
edge from parent
node[above] {$q_2$}
% node[below] {8}
}
edge from parent
% node[above] {}
node[below] {$1-q_1$}
}
child {
node[bag] {\underline{$B_i'=1$}}
child {
node[end, label=right:
{$B_i''=0$}] {}
edge from parent
% node[above] {13}
node[below] {$1-p_2$}
}
child {
node[end, label=right:
{\underline{$B_i''=1$}}] {}
edge from parent
node[above] {$p_2$}
% node[below] {17}
}
edge from parent
node[above] {$q_1$}
% node[below] {19}
};
\end{tikzpicture}
\caption{Probability trees for two rounds of sanitization using UE (L-UE).} 
\label{fig:tree_l_ue}
\end{figure}

Visually, Fig.~\ref{fig:tree_l_ue} illustrates the probability tree of the L-UE protocol. \textbf{One natural question emerges: how to select the parameters $\{p_1,q_1,p_2,q_2\}$ in order to optimize the utility of this L-UE protocol?} One can see $RR_1[UE]$ as a \textit{permanent} sanitization and $RR_2[UE]$ as a `small' perturbation to avoid averaging attacks and keep privacy over time. 

Based on SUE and OUE, we are then left with four options: two popular solutions that strictly use only OUE or SUE parameters in both sanitization steps and two proposed settings that combine both OUE and SUE. These four L-UE protocols are summarized below: 

\begin{enumerate}[I]
    \item both sanitizations with OUE (L-OUE);
    \item both sanitizations with SUE (L-SUE);
    \item starting with OUE and then with SUE (L-OSUE);
    \item starting with SUE and then with OUE (L-SOUE);
\end{enumerate}

\noindent in which L-SUE is the well-known Basic-RAPPOR protocol~\cite{rappor}, L-OUE is the state-of-the-art OUE protocol~\cite{tianhao2017} with memoization, and both L-OSUE and L-SOUE are proposed in this paper. 

As presented in~\cite{tianhao2017}, the OUE variance in Eq.~\eqref{eq:var_oue} is smaller than the SUE variance in Eq.~\eqref{eq:var_sue} and, therefore, the former can provide higher utility than the latter for $RR_1$. On the other hand, we argue that OUE might be too strict for $RR_2$ since the parameter $p_2=1/2$ is constant. Thus, we hypothesize that option III (i.e., L-OSUE) is the most suitable one. Without losing generality, \textbf{the following analyses are done only for L-OSUE}, which can be easily extended to any of the other combinations. 

In the first round of sanitization, $RR_1$, our solution L-OSUE applies OUE with $p_1=\Pr[ B_i^{'} =1 | B_i=1 ] =\frac{1}{2}$ and $q_1=\Pr[ B_i^{'} =1 | B_i=0 ] =\frac{1}{e^{\epsilon_{\infty}}+1}$ (\underline{underlined} in the middle of Fig.~\ref{fig:tree_l_ue}). As discussed in Section~\ref{subsub:UE}, this \textit{permanent} memoization satisfies $\epsilon_{\infty}$-LDP since $\frac{p_1(1-q_1)}{(1-p_1)q_1}=e^{\epsilon_{\infty}}$, which is the upper bound. 

Following the same development as for L-GRR, on the other hand, with a single collection of data, the attacker's knowledge of $B=UE(v)$ comes only from $B''$, which is generated using two randomization steps with OUE and SUE, respectively. This provides a higher level of privacy protection~\cite{rappor}. From Fig.~\ref{fig:tree_l_ue}, we can obtain the following conditional probabilities according to each bit $i \in [1,k_j]$:

\begin{equation*}
    \begin{gathered}
    \Pr[ B_i'' | B_i ] = \\
    \begin{cases}
        \Pr[ B_i''= 1| B_i = 1] = p_1 p_2 + (1 - p_1) q_2 &\\
        \Pr[ B_i''= 0| B_i = 1] = p_1 (1 - p_2) + (1 - p_1) (1 - q_2) \\
        \Pr[ B_i''= 1| B_i = 0] = q_1 p_2 + (1 - q_1) q_2 \\
        \Pr[ B_i''= 0| B_i = 0] = q_1 (1 - p_2) + (1 - q_1) (1 - q_2) \\
    \end{cases}
    \end{gathered}
\end{equation*}

Let $p_s=\Pr[ B_i'' =1 | B_i =1]$ and $q_s = \Pr[ B_i'' =1 | B_i =0 ]$ (\underline{underlined} in far right of Fig.~\ref{fig:tree_l_ue}), with the second round of sanitization, $RR_2[SUE]$, our proposed L-OSUE protocol satisfies $\epsilon_1$-LDP since $\frac{p_s(1-q_s)}{(1-p_s)q_s}=e^{\epsilon_{1}}$. Notice that $\epsilon_{1}$ corresponds to a single report (lower bound) and its extension to infinity reports is limited by $\epsilon_{\infty}$ (upper bound) since $RR_2[SUE]$ uses as input the output of $RR_1[OUE]$. More specifically, the calculus of $\epsilon_1$ for L-OSUE (or L-UE protocols in general) is:

\begin{equation}\label{eq:e1_ue}
    \epsilon_{1} = \ln{  \left (  \frac{\left(p_{1} p_{2} - q_{2} \left(p_{1} - 1\right)\right) \left(p_{2} q_{1} - q_{2} \left(q_{1} - 1\right) - 1\right)}{\left(p_{2} q_{1} - q_{2} \left(q_{1} - 1\right)\right) \left(p_{1} p_{2} - q_{2} \left(p_{1} - 1\right) - 1\right)}  \right)} 
\end{equation}

\noindent in which, for L-OSUE, we have $p_1=\frac{1}{2}$, $q_1=\frac{1}{e^{\epsilon_{\infty}}+1}$, and both $p_2$ and $q_2$ are symmetric ($p_2+q_2 = 1$) and selectable according to $\epsilon_{\infty}$ and $\epsilon_1$, calculated as: 

\begin{equation}  \label{eq:p2_losue}
    \begin{gathered}
    p_2 = \frac{1 - e^{\epsilon_{1} + \epsilon_{\infty}}}{e^{\epsilon_{1}} - e^{\epsilon_{\infty}} - e^{\epsilon_{1} + \epsilon_{\infty}} + 1} \\
    q_2 = 1 - p_2 
    \end{gathered}
\end{equation}

Similarly, the estimated frequency $\hat{f}_L(v_i)$ that a value $v_i$ occurs for $i \in [1,k_j]$ is calculated using Eq.~\eqref{eq:est_longitudinal}. Lastly, one can calculate the L-OSUE (or L-UE protocols in general) approximate variance by replacing the resulting $p_1,q_1,p_2,q_2$ parameters into Eq.~\eqref{var:aprox_longitudinal}.

\subsection{Numerical evaluation of L-GRR and L-UE protocols}\label{sub:analysis_long}

In this subsection, we evaluate numerically the approximate variance of all developed longitudinal protocols, namely, L-GRR, and the four UE-based options, namely, L-OUE, L-SUE, L-OSUE, and L-SOUE, respectively. As aforementioned, once both $\epsilon_{\infty}$ and $\epsilon_1$ privacy guarantees are defined, one can obtain parameters $p_1$ and $q_1$ depending on $\epsilon_{\infty}$, and parameters $p_2$ and $q_2$ depending on both $\epsilon_{\infty}$ and $\epsilon_1$ (and the domain size $k_j$ for L-GRR), as given in Eq.~\eqref{eq:p2_lgrr} for L-GRR and in Eq.~\eqref{eq:p2_losue} for L-OSUE. 

Next, once the parameters $\{p_1,q_1,p_2,q_2\}$ are computed, one can calculate the approximate variance with Eq.~\eqref{var:aprox_longitudinal} for each protocol. In other words, following our proposal, one has to set both the upper ($\epsilon_{\infty}$) and lower ($\epsilon_1$) bounds of the privacy guarantees. For example, let $\epsilon_{\infty} = 2$, one might want the first $\epsilon_1$-LDP report to have high privacy such as $\epsilon_1=0.1$, i.e., $\epsilon_1=0.05\epsilon_{\infty}$ (\textbf{we will use this percentage notation to set up the privacy guarantees}).

Table~\ref{tab:analysis_var} exhibits the numerical values of the approximate variance using Eq.~\eqref{var:aprox_longitudinal} for all longitudinal protocols with $n=10000$, $\epsilon_{\infty}=[0.5, 1.0, 2.0, 4.0]$ (as in~\cite{tianhao2017}), and $\epsilon_1 = \{0.6\epsilon_{\infty},0.5\epsilon_{\infty},0.4\epsilon_{\infty},0.3\epsilon_{\infty},0.2\epsilon_{\infty},0.1\epsilon_{\infty}\}$. For values of $\epsilon_1$ higher than $0.6\epsilon_{\infty}$, neither L-OUE nor L-SOUE could satisfy some values of $\epsilon_1$ because of the constant $p_2=1/2$ in $RR_2$. However, it is not desirable to have higher values of $\epsilon_1$ and, thus, we do not consider values above $0.6\epsilon_{\infty}$ in our analysis. Besides, Table~\ref{tab:analysis_var_non_long} exhibits the numerical values for the non-longitudinal GRR, OUE, and SUE protocols, which allow evaluating how utility degrades with a second step of sanitization. 

\setlength{\tabcolsep}{6pt}
\renewcommand{\arraystretch}{1.2}
\begin{table*}[!ht]
    \scriptsize
    \centering
    \begin{tabular}{c c| c| c| c| c| c| c| c}\hline
    \multicolumn{2}{c|}{\multirow{2}{*}{Privacy Guarantees}} &  \multicolumn{3}{c|}{L-GRR} &  \multicolumn{4}{c}{L-UE}  \\ \cline{3-9}
    & & $k_j=2$ & $k_j=32$ & $k_j=2^{10}$ & L-OSUE & L-SUE & L-SOUE & L-OUE\\ \hline
     
     \multirow{4}{*}{$\epsilon_{1}=0.6\epsilon_{\infty}$} &$\epsilon_{\infty}=0.5,\epsilon_{1}=0.30$ & 0.001103 &     0.980969 &       26706 &  0.004411 &  0.004436 &  0.005306 &  0.005549 \\
     & $\epsilon_{\infty}=1.0,\epsilon_{1}=0.60$ &    0.000270 &     0.125036 &        3153 &  0.001078 &  0.001103 &  0.001234 &  0.001347 \\
     & $\epsilon_{\infty}=2.0,\epsilon_{1}=1.20$ & 0.000062 &     0.006327 &         117 &  0.000247 &  0.000270 &  0.000264 &  0.000310 \\    
     & $\epsilon_{\infty}=4.0,\epsilon_{1}=2.40$ &    0.000011 &     0.000078 &           0.25903 &  0.000044 &  0.000062 &  0.000045 &  0.000057 \\ \hline
     
     \multirow{4}{*}{$\epsilon_{1}=0.5\epsilon_{\infty}$} &$\epsilon_{\infty}=0.5,\epsilon_{1}=0.25$ & 0.001592 &     2.088372 &       60218 &  0.006367 &  0.006392 &  0.007336 &  0.007611 \\ %\cline{2-8}
     &$\epsilon_{\infty}=1.0,\epsilon_{1}=0.50$ & \textbf{0.000392} &     0.268074 &        7198 &  \textbf{0.001567} &  \textbf{0.001592} &  0.001740 &  0.001872   \\ %\cline{2-8}
     &$\epsilon_{\infty}=2.0,\epsilon_{1}=1.00$ & \textbf{0.000092} &     0.013926 &         281 &  \textbf{0.000368} &  \textbf{0.000392} &  0.000389 &  0.000447   \\ %\cline{2-8}
     &$\epsilon_{\infty}=4.0,\epsilon_{1}=2.00$ & \textbf{0.000018} &     0.000188 &           0.74088 &  \textbf{0.000072} &  \textbf{0.000092} &  0.000073 &  0.000092   \\ \hline
     
     \multirow{4}{*}{$\epsilon_{1}=0.4\epsilon_{\infty}$} &$\epsilon_{\infty}=0.5,\epsilon_{1}=0.20$ &  0.002492 &     4.530779 &      135874 &  0.009967 &  0.009992 &  0.011012 &  0.011324 \\
     &$\epsilon_{\infty}=1.0,\epsilon_{1}=0.40$ &  0.000617 &     0.586823 &       16443 &  0.002467 &  0.002492 &  0.002658 &  0.002812\\ 
     &$\epsilon_{\infty}=2.0,\epsilon_{1}=0.80$ &  0.000148 &     0.031552 &         673 &  0.000593 &  0.000617 &  0.000617 &  0.000690  \\ 
     &$\epsilon_{\infty}=4.0,\epsilon_{1}=1.60$ &  0.000032 &     0.000484 &           2.12772 &  0.000127 &  0.000148 &  0.000128 &  0.000156  \\ \hline
    
     \multirow{4}{*}{$\epsilon_{1}=0.3\epsilon_{\infty}$} &$\epsilon_{\infty}=0.5,\epsilon_{1}=0.15$ & 0.004436 &    10 &      329836 &  0.017744 &  0.017769 &  0.018863 &  0.019214 \\
     &$\epsilon_{\infty}=1.0,\epsilon_{1}=0.30$ & 0.001103 &     1.398568 &       40412 &  0.004411 &  0.004436 &  0.004620 &  0.004799 \\
     &$\epsilon_{\infty}=1.0,\epsilon_{1}=0.60$ & 0.000270 &     0.078202 &        1737 &  0.001078 &  0.001103 &  0.001106 &  0.001198 \\
     &$\epsilon_{\infty}=2.0,\epsilon_{1}=1.20$ & 0.000062 &     0.001389 &           6 &  0.000247 &  0.000270 &  0.000248 &  0.000291 \\ \hline
          
     \multirow{4}{*}{$\epsilon_{1}=0.2\epsilon_{\infty}$} &$\epsilon_{\infty}=0.5,\epsilon_{1}=0.10$ &  0.009992 &    30 &      972656 &  0.039967 &  0.039992 &  0.041148 &  0.041536 \\
     &$\epsilon_{\infty}=1.0,\epsilon_{1}=0.20$ & 0.002492 &     4.080052 &      120651 &  0.009967 &  0.009992 &  0.010190 &  0.010394  \\ %\cline{2-8}
     &$\epsilon_{\infty}=2.0,\epsilon_{1}=0.40$ & 0.000617 &     0.237925 &        5443 &  0.002467 &  0.002492 &  0.002498 &  0.002610   \\ %\cline{2-8}
     &$\epsilon_{\infty}=4.0,\epsilon_{1}=0.80$ & 0.000148 &     0.004939 &          24 &  0.000593 &  0.000617 &  0.000595 &  0.000659   \\ \hline
     
     \multirow{4}{*}{$\epsilon_{1}=0.1\epsilon_{\infty}$} & $\epsilon_{\infty}=0.5,\epsilon_{1}=0.05$ & 0.039992 &   154 &     4941829 &  0.159967 &  0.159992 &  0.161191 &  0.161608 \\ %\cline{2-8}
     & $\epsilon_{\infty}=1.0,\epsilon_{1}=0.10$ & 0.009992 &    20 &      620584 &  0.039967 &  0.039992 &  0.040201 &  0.040424 \\ %\cline{2-8}
     & $\epsilon_{\infty}=2.0,\epsilon_{1}=0.20$ & 0.002492 &     1.255550 &       29356 &  0.009967 &  0.009992 &  0.010000 &  0.010130  \\ %\cline{2-8}
     & $\epsilon_{\infty}=4.0,\epsilon_{1}=0.40$ & 0.000617 &     0.030494 &         156 &  0.002467 &  0.002492 &  0.002469 &  0.002560  \\ \hline
    \end{tabular}
    \caption{Numerical values of Eq.~\eqref{var:aprox_longitudinal} (i.e., $Var^*[\hat{f}_L(v_i)]$) for L-GRR and L-UE protocols with different $\epsilon_{\infty}$ and $\epsilon_1$ privacy guarantees, following $\epsilon_1 = \{0.6\epsilon_{\infty},0.5\epsilon_{\infty},0.4\epsilon_{\infty},0.3\epsilon_{\infty},0.2\epsilon_{\infty},0.1\epsilon_{\infty}\}$, respectively.}
    \label{tab:analysis_var}
\end{table*}

\setlength{\tabcolsep}{2.4pt}
\renewcommand{\arraystretch}{1.2}
\begin{table}[!ht]
    \scriptsize
    \centering
    \begin{tabular}{c | c| c| c| c| c}\hline
    $\epsilon_{\infty}$ &GRR($k_j=2$)  &GRR($k_j=32$)  &GRR($k_j=2^{10}$)  &OUE  &SUE \\ \hline
    $\epsilon_{\infty}=0.5$    &\textbf{0.000392} &     0.007520 &           0.243240 &  \textbf{0.001567} &  \textbf{0.001592} \\ \hline
    $\epsilon_{\infty}=1.0$    &\textbf{0.000092} &     0.001108 &           0.034707 &  \textbf{0.000368} &  \textbf{0.000392} \\ \hline
    $\epsilon_{\infty}=2.0$    &\textbf{0.000018} &     0.000092 &           0.002522 &  \textbf{0.000072} &  \textbf{0.000092} \\ \hline
    $\epsilon_{\infty}=4.0$    &0.000002 &     0.000003 &           0.000037 &  0.000008 &  0.000018  \\ \hline
    \end{tabular}
    \caption{Numerical values of $Var^*[\hat{f}(v_i)]$ for the non-longitudinal GRR, OUE, and SUE protocols with different $\epsilon_{\infty}$ privacy guarantees.}
    \label{tab:analysis_var_non_long}
\end{table}

\textbf{From Table~\ref{tab:analysis_var}, one can notice that L-GRR presents the smallest variance values for binary attributes (i.e., when $k_j=2$).} On the other hand, L-GRR is also most sensitive to changes in privacy parameters $\epsilon_{\infty}$ and $\epsilon_1$ when $k_j$ is large, which shows a much higher variance than when using a non-longitudinal GRR, as shown in Table~\ref{tab:analysis_var_non_long}. Similar to the non-longitudinal GRR, this increase in the variance is due to the number of values $k_j$, which decreases the probability $p$ of reporting the true value. With two rounds of sanitization, it further deteriorates the accuracy of the L-GRR protocol that gets extremely high values, e.g., see L-GRR$(k_j=2^{10})$. Interestingly, when $k_j=2$ in Table~\ref{tab:analysis_var}, the variance of L-GRR with $\epsilon_1=0.5\epsilon_{\infty}$ is a lagged version of the variance values given by the non-longitudinal GRR in Table~\ref{tab:analysis_var_non_long}. This effect is also observed for both the L-SUE (cf. SUE in Table~\ref{tab:analysis_var_non_long}) and L-OSUE (cf. OUE in Table~\ref{tab:analysis_var_non_long}) protocols, which use symmetric probabilities on $RR_2$ (i.e., $p_2+q_2=1$). We highlight these values in \textbf{bold font}. However, for L-GRR, this is not true for other values of $k_j$, the further analysis of which is beyond the scope of this paper.

\begin{figure}[!ht]
    \centering
    \includegraphics[width=1\linewidth]{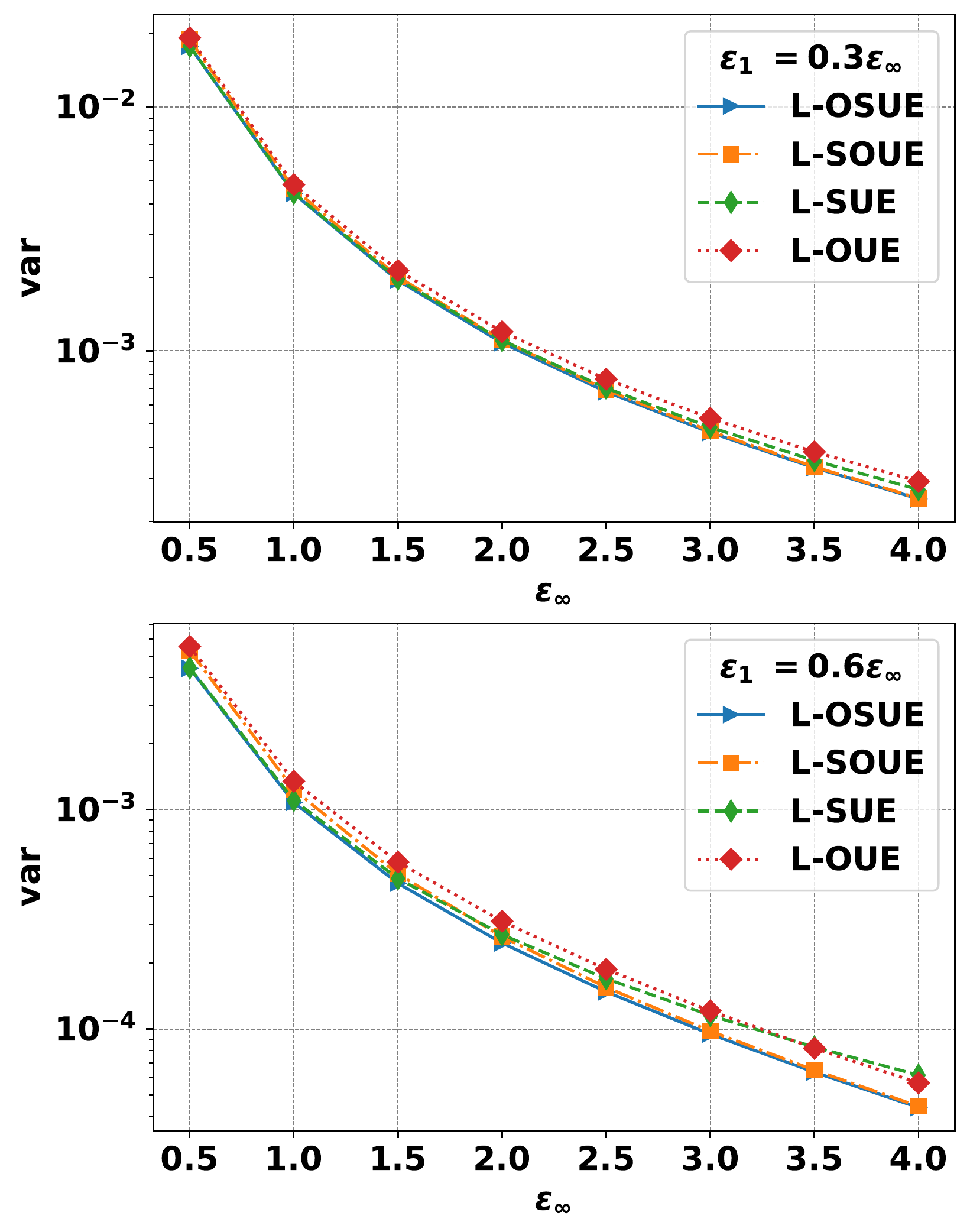}
    \caption{Numerical values of $Var^*[\hat{f}_L(v_i)]$ for L-UE protocols with $\epsilon_{1}=0.3\cdot \epsilon_{\infty}$ (plot on the top) and with $\epsilon_{1}=0.6\cdot \epsilon_{\infty}$ (plot on the bottom).}
    \label{fig:analysis_var}
\end{figure}

On the other hand, the L-UE protocols avoid having a variance that depends on $k_j$ by encoding the value into the unary representation, which results in a constant variance regardless of the size of the attribute. To complement the results of Table~\ref{tab:analysis_var}, Fig.~\ref{fig:analysis_var} illustrates the numerical values of the approximate variance for the L-UE protocols with $\epsilon_1=\{0.3\epsilon_{\infty}, 0.6\epsilon_{\infty}\}$. With the four options I-IV analyzed, on the high privacy regimes, L-OSUE and L-SUE have similar performance while \textit{always} favoring the proposed L-OSUE. On lower privacy regimes, our proposed protocols L-SOUE and L-OSUE have similar performance, which outperform both the L-OUE and L-SUE protocols. As shown in our experiments, the L-OUE protocol has the worst performance among the four options analyzed, with the exception of high values for $\epsilon_{\infty}$ (see the plot on the bottom of Fig.~\ref{fig:analysis_var}), when it has performance superior or similar to that of L-SUE. Indeed, for L-OUE, selecting $p_2=1/2$ for the second sanitization step is too strict, which results in higher variance values. Therefore, by comparing the approximate variances, \textbf{the best option for L-UE protocols, in terms of utility, is to start with OUE and then with SUE as we propose in this paper, i.e., L-OSUE.}

\subsection{The \textit{ALLOMFREE} algorithm}\label{sub:allomfree}

Let $A=\{A_1,A_2,...,A_d\}$ be a set of $d$ attributes with the domain size $\textbf{k}=[k_1,k_2,...,k_d]$, $\mathbb{A}=\{\textit{L-GRR},\textit{L-OSUE}\}$ be a set of optimal longitudinal LDP protocols, and $\epsilon_{\infty}$ and $\epsilon_1$ be the longitudinal and \textit{single-report} privacy guarantees, respectively. Each user $u_i$, for $1 \leq i \leq n$, holds a tuple $\textbf{v}^{(i)}=(v^{(i)}_{1},v^{(i)}_{2},...,v^{(i)}_{d})$, i.e., a private value per attribute. From now on, we will simply omit the index notation $\textbf{v}^{(i)}$ and use $\textbf{v}$ in the analysis as we focus on one arbitrary user $u_i$ here. For each attribute $j \in [1,d]$ (we slightly abuse the notation and use $j$ for $A_j$) at time $t \in [1,\tau]$, the aggregator aims to estimate the frequencies of each value $v \in A_j$. 

\noindent \textbf{Client-Side.} In a multidimensional setting with different domain sizes for each attribute, a dynamic selection of longitudinal LDP protocols is preferred. As mentioned in Section~\ref{sec:multidimensional}, we propose that each user randomly sample $r=Uniform(1,2,...,d)$ to select a single attribute $A_r$. Given $k_r$ (the domain size), $\epsilon_{\infty}$, and $\epsilon_1$, one calculates the parameters $fp_{L-GRR}=\{p_1,q_1,p_2,q_2\}$ and $fp_{L-OSUE}=\{p_1,q_1,p_2,q_2\}$, for L-GRR and L-OSUE, respectively (cf. Eq.~\eqref{eq:p2_lgrr} and Eq.~\eqref{eq:p2_losue}). Next, with $fp_{L-GRR}$ and $fp_{L-OSUE}$, one calculates the approximate variances $Var^*[\hat{f}_{L_{(\textit{L-GRR})}}]$ for L-GRR and $Var^*[\hat{f}_{L_{(\textit{L-OSUE})}}]$ for L-OSUE with Eq.~\eqref{var:aprox_longitudinal}. Lastly, to select L-GRR as the local randomizer, we are then left to evaluate if $Var^*[\hat{f}_{L_{(\textit{L-GRR})}}] \leq Var^*[\hat{f}_{L_{(\textit{L-OSUE})}}]$. Therefore, the first round of sanitization ensures a \textit{permanent memoization} $B'$ that is always used for the second round of sanitization to generate $B''$ each time $t \in [1,\tau]$ the user will report the real value $B$. We call our solution \underline{A}daptive \underline{L}DP for \underline{LO}ngitudinal and \underline{M}ultidimensional \underline{FRE}quency \underline{E}stimates (ALLOMFREE), which is summarized in Algorithm~\ref{alg:allomfree} as a pseudocode. 

\begin{algorithm*}[!ht]
\caption{User-side algorithm of ALLOMFREE.}
\label{alg:allomfree}
\begin{algorithmic}[1]
\State \textbf{Input :} $\textbf{v} = [v_1,v_2,..., v_d]$, $\textbf{k}=[k_1,k_2,...,k_d]$, $\mathbb{A}=\{\textit{L-GRR},\textit{L-OSUE}\}$, $\epsilon_{\infty}$, $\epsilon_1$, number of reports $\tau$. 

\State $r \gets Uniform(\{1,2,...,d \})$ \Comment{Select attribute only once}

\State $B \gets \texttt{Encode}(v_r)$ \Comment{Encode (if needed)}

\State $fp_{L-GRR} \gets p_1=\frac{e^{\epsilon_{\infty}}}{e^{\epsilon_{\infty}}+k_r-1},q_1=\frac{1-p_1}{k_r-1}, p_2 = \frac{e^{\epsilon_{1} + \epsilon_{\infty}} - 1}{- k_r e^{\epsilon_{1}} + \left(k_r - 1\right) e^{\epsilon_{\infty}} + e^{\epsilon_{1}} + e^{\epsilon_{1} + \epsilon_{\infty}} - 1}, q_2  = \frac{1-p_2}{k_r - 1}$ \Comment{Get $p_2$ and $q_2$ with Eq.~\eqref{eq:e1_grr}}

\State $fp_{L-OSUE} \gets p_1=\frac{1}{2}, q_1=\frac{1}{e^{\epsilon_{\infty}}+1}, p_2 = \frac{1 - e^{\epsilon_{1} + \epsilon_{\infty}}}{e^{\epsilon_{1}} - e^{\epsilon_{\infty}} - e^{\epsilon_{1} + \epsilon_{\infty}} + 1}, q_2 = 1 - p_2$ \Comment{Get $p_2$ and $q_2$ with Eq.~\eqref{eq:e1_ue}}

\State \textbf{if}  $Var^*[\hat{f}_{L_{(\textit{L-GRR})}}](fp_{L-GRR}) \leq Var^*[\hat{f}_{L_{(\textit{L-OSUE})}}](fp_{L-OSUE})$  : \Comment{Check variances with Eq.~\eqref{var:aprox_longitudinal}}

\State  \hskip1em $\mathcal{A} \gets \textrm{L-GRR}$ \Comment{Select L-GRR as local randomizer}

\State \textbf{else}

\State  \hskip1em $\mathcal{A} \gets \textrm{L-OSUE}$ \Comment{Select L-OSUE as local randomizer}

\State $B' \gets \mathcal{A}(B, p_1, q_1, k_r)$ \Comment{First round of sanitization (permanent memoization)}

\State \textbf{for} $t \in [1,\tau]$ \textbf{do}

\State  \hskip1em $B''= \mathcal{A}(B', p_2, q_2, k_r)$ \Comment{Second round of sanitization}

\State \textbf{end for}

\State  \textbf{send :} $(t,\langle r, B''\rangle)$ for $t \in [1,\tau]$ 

\end{algorithmic}
\end{algorithm*}

The intuition of ALLOMFREE is as follows. By requiring each user to submit only 1 attribute with the whole privacy budget, it reduces both the variance incurred as well as the communication cost. Also, since we develop the calculus of the approximate variance in Eq.~\eqref{var:aprox_longitudinal} for the proposed longitudinal protocols (L-GRR and L-OSUE), ALLOMFREE can adaptively select the protocol with a smaller variance value to optimize the data utility. Therefore, ALLOMFREE utilizes optimal solutions for both multidimensional and longitudinal data collection settings developed in Sections~\ref{sec:multidimensional} and~\ref{sec:longitudinal} of this paper, respectively.

\noindent \textbf{Server-Side.} On the server-side, for each attribute $j\in[1,d]$ at time $t \in [1,\tau]$, the estimated frequency $\hat{f}_L(v_i)$ that a value $v_i$ occurs for $i \in [1,k_j]$ is calculated using Eq.~\eqref{eq:est_longitudinal}.

\noindent \textbf{Privacy analysis.} On the one hand, according to the analysis in subsections~\ref{subsub:l_de} and \ref{subsub:l_ue}, Alg.~\ref{alg:allomfree} satisfies $\epsilon$-LDP with upper $\epsilon_{\infty}$ (infinity reports) and lower $\epsilon_1$ (a single report) bounds as it uses either L-GRR or L-OSUE to sanitize a single attribute per user. \textbf{Notice that, to ensure the users' privacy over time and to avoid the sequential composition theorem~\cite{dwork2014algorithmic}, each user must always report the same unique attribute $A_r$}. In addition, the privacy of a user decreases gracefully according to the number of LDP reports $t \leq \tau $ that an adversary has gained access to, which is calculated as~\cite{Naor2020,erlingsson2020encode}: 

\begin{equation}\label{eq:eps_long}
\epsilon_t=\ln{\left (\frac{e^{\epsilon_{{\infty}}+t\epsilon_{1}} + 1}{e^{\epsilon_{{\infty}}}+e^{t\epsilon_{1}}} \right)} \leq \min \{ \epsilon_{{\infty}}, t\epsilon_{1}\}    
\end{equation}

\noindent \textbf{Limitations.} Similar to other sampling-based methods for collecting multidimensional data under LDP~\cite{Duchi2018,xiao2,wang2019,Wang2021_b}, our ALLOMFREE algorithm also entails a \textit{sampling error}, which is due to observing a sample instead of the entire population. In addition, concerning the privacy guarantees, the memoization step of ALLOMFREE is certainly effective for longitudinal privacy in the cases where the true client's data does not vary (static) or vary very slowly or in an uncorrelated manner~\cite{rappor}. In many application scenarios, gender, age range, nationality, and other demographic data are generally static or hardly ever vary. On the other hand, for dynamic attributes such as the location or the time spent in the application, this is not the case. Therefore, for each different value, a new memoized value would be generated, thus accumulating the privacy budget $\epsilon_{\infty}$ by the sequential composition theorem~\cite{dwork2014algorithmic}.

\section{Experimental results} \label{sec:results_discussion}

In this section, we present the setup of our experiments and the results with real-world data.

\subsection{Setup of experiments} \label{sub:setup}

The main goal of our experiments is to evaluate the proposed longitudinal LDP protocols on multidimensional frequency estimates a single time, i.e., satisfying $\epsilon_1$-LDP (as in~\cite{rappor,Vidal2020,Kim2018}, for example).

\noindent \textbf{Environment.} All algorithms are implemented in Python 3.8.8 with NumPy 1.19.5 and Numba 0.53.1 libraries. The codes we develop and use for all experiments are available in a Github repository\footnote{\url{https://github.com/hharcolezi/ldp-protocols-mobility-cdrs}}. In all experiments, we report average results over 100 runs as LDP algorithms are randomized.

\noindent \textbf{Methods evaluated.} We consider for evaluation the following solutions and protocols: 

\begin{itemize}
    
    \item Solution \textit{Smp} (cf. Section~\ref{sec:multidimensional}), which randomly samples a single attribute to be sent with the whole privacy budget. We will experiment with the state-of-the-art protocols, namely, L-SUE and L-OUE, and with our extended protocols L-OSUE and L-SOUE; %That is, L-SUE[Smp] and L-OUE[Smp];
    
    \item Our ALLOMFREE solution (cf. Alg.~\ref{alg:allomfree}), which also randomly samples a single attribute to be sent with the whole privacy budget but adaptively select the optimal protocol, i.e., either L-GRR or L-OSUE.

\end{itemize}

\noindent \textbf{Experimental evaluation and metrics.} We vary the longitudinal privacy parameter in the range $\epsilon_{\infty}=[0.5, 1, ..., 3.5, 4]$ with $\epsilon_1 = [0.3\epsilon_{\infty}, 0.6 \epsilon_{\infty}]$ to compare our experimental results with numerical ones from subsection~\ref{sub:analysis_long}. Notice that this range of privacy guarantees is commonly used in the literature for multidimensional data (e.g., in~\cite{wang2019} the range is $\epsilon=[0.5,...,4]$ and in~\cite{Wang2021_b} the range is $\epsilon=[0.1,...,10]$). 

To evaluate our results, we use the MSE metric averaged per the number of attributes $d$ \textbf{in a single data collection $\tau=1$, i.e., with $\epsilon_1$-LDP}. Thus, for each attribute $j$, we compute for each value $v_i \in A_j$ the estimated frequency $\hat{f}(v_i)$ and the real one $f(v_i)$ and calculate their differences. More precisely,

\begin{equation*}
    MSE_{avg} = \frac{1}{\tau} \sum_{t \in [1,\tau]} \frac{1}{d} \sum_{j \in [1,d]} \frac{1}{|A_j|} \sum_{v \in A_j}(f(v_i) - \hat{f}(v_i) )^2 
\end{equation*}

\noindent \textbf{Datasets.} For the ease of reproducibility, we conduct our experiments on four multidimensional open datasets. 

\begin{itemize}
    \item \textit{Nursery.} A dataset from the UCI machine learning repository~\cite{uci} with $d=9$ categorical attributes and $n=12960$ samples. The domain size of each attribute is $\textbf{k}=[3, 5, 4, 4, 3, 2, 3, 3, 5]$, respectively. 
    
    \item \textit{Adult.} A dataset from the UCI machine learning repository~\cite{uci} with $d=9$ categorical attributes and $n=45222$ samples after cleaning the data. The domain size of each attribute is $\textbf{k}=[7, 16, 7, 14, 6, 5, 2, 41, 2]$, respectively. 
    
    \item \textit{MS-FIMU.} An open dataset from~\cite{ms_fimu} with $d=6$ categorical attributes and $n=88935$ samples. The domain size of each attribute is $\textbf{k}=[3, 3, 8, 12, 37, 11]$, respectively.
    
    \item \textit{Census-Income.} A dataset from the UCI machine learning repository~\cite{uci} with $d=33$ categorical attributes and $n=299285$ samples. The domain size of each attribute is \begin{math}\textbf{k}=[9, 52, 47, 17,  3,  ..., 43, 43, 43,  5,  3,  3,  3,  2]\end{math}, respectively. 

\end{itemize}

\subsection{Results} \label{sub:results}

\begin{figure}
    \centering
    \includegraphics[width=1\linewidth]{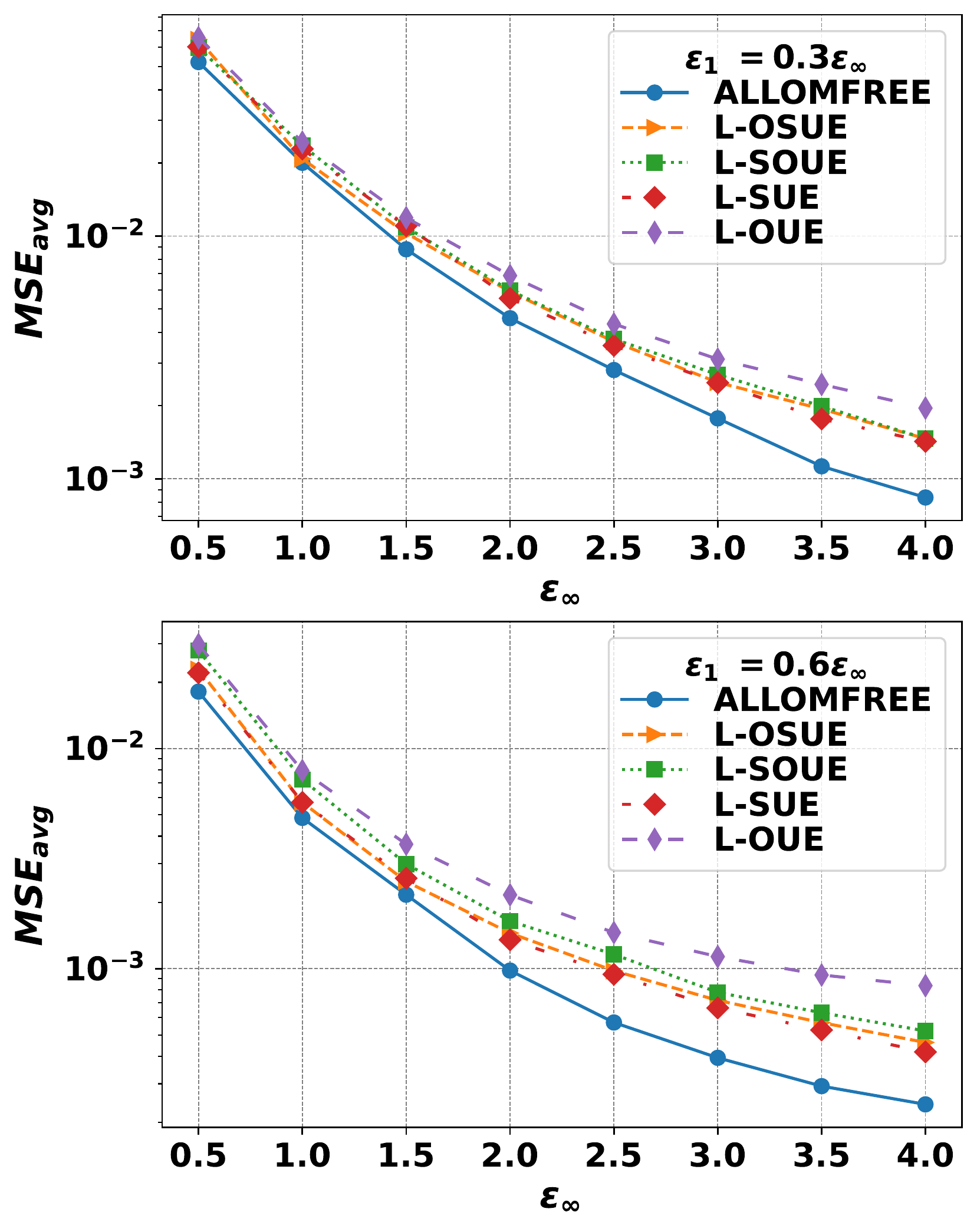}\\
    \caption{Averaged MSE varying $\epsilon_{\infty}$ with $\epsilon_1 = 0.3\epsilon_{\infty}$ (plot on the top) and with $\epsilon_1 = 0.6\epsilon_{\infty}$ (plot on the bottom) on the \textit{Nursery} dataset.}% with $n=12960$, $d=9$, and domain size $\textbf{k}=[3, 5, 4, 4, 3, 2, 3, 3, 5]$
    \label{fig:results_nursery}
    \includegraphics[width=1\linewidth]{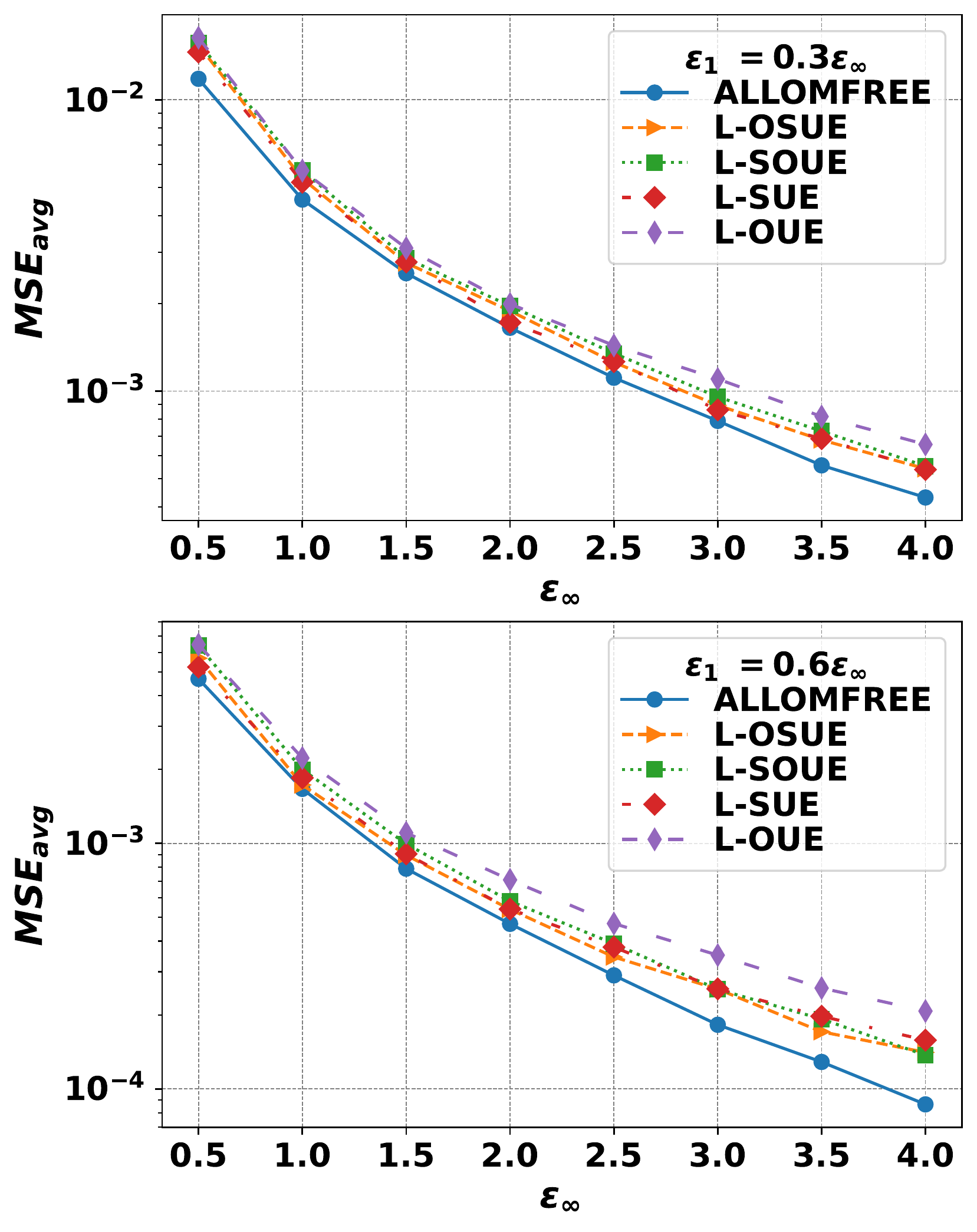}\\
    \caption{Averaged MSE varying $\epsilon_{\infty}$ with $\epsilon_1 = 0.3\epsilon_{\infty}$ (plot on the top) and with $\epsilon_1 = 0.6\epsilon_{\infty}$ (plot on the bottom) on the \textit{Adult} dataset.}% with $n=45222$, $d=9$, and domain size $\textbf{k}=[7, 16, 7, 14, 6, 5, 2, 41, 2]$
    \label{fig:results_adults}
\end{figure}

\begin{figure}
    \centering
    \includegraphics[width=1\linewidth]{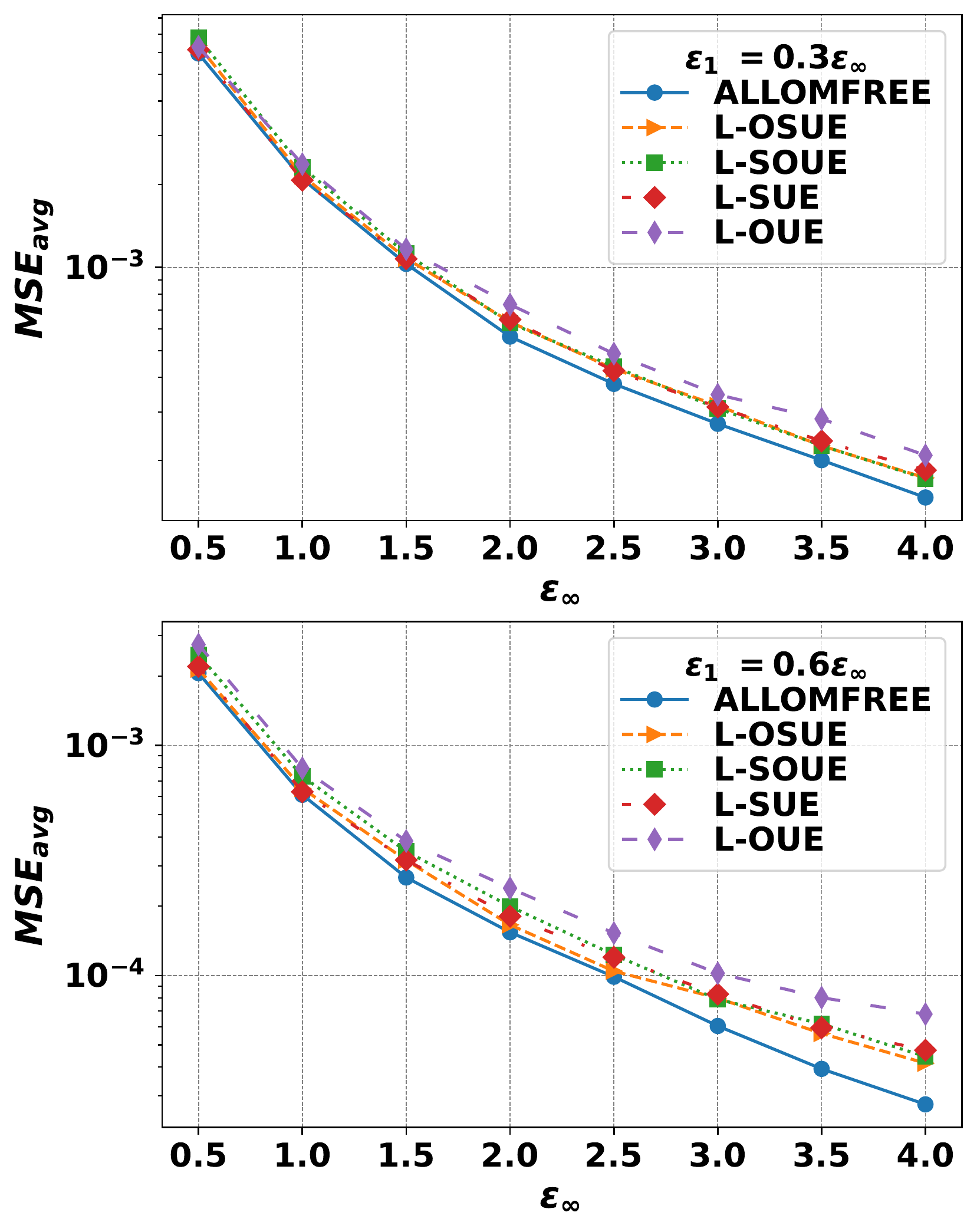}\\
    \caption{Averaged MSE varying $\epsilon_{\infty}$ with $\epsilon_1 = 0.3\epsilon_{\infty}$ (plot on the top) and with $\epsilon_1 = 0.6\epsilon_{\infty}$ (plot on the bottom) on the \textit{MS-FIMU} dataset.}% with $n=88935$, $d=6$, and domain size $\textbf{k}=[3, 3, 8, 12, 37, 11]$
    \label{fig:results_vhs}
    \includegraphics[width=1\linewidth]{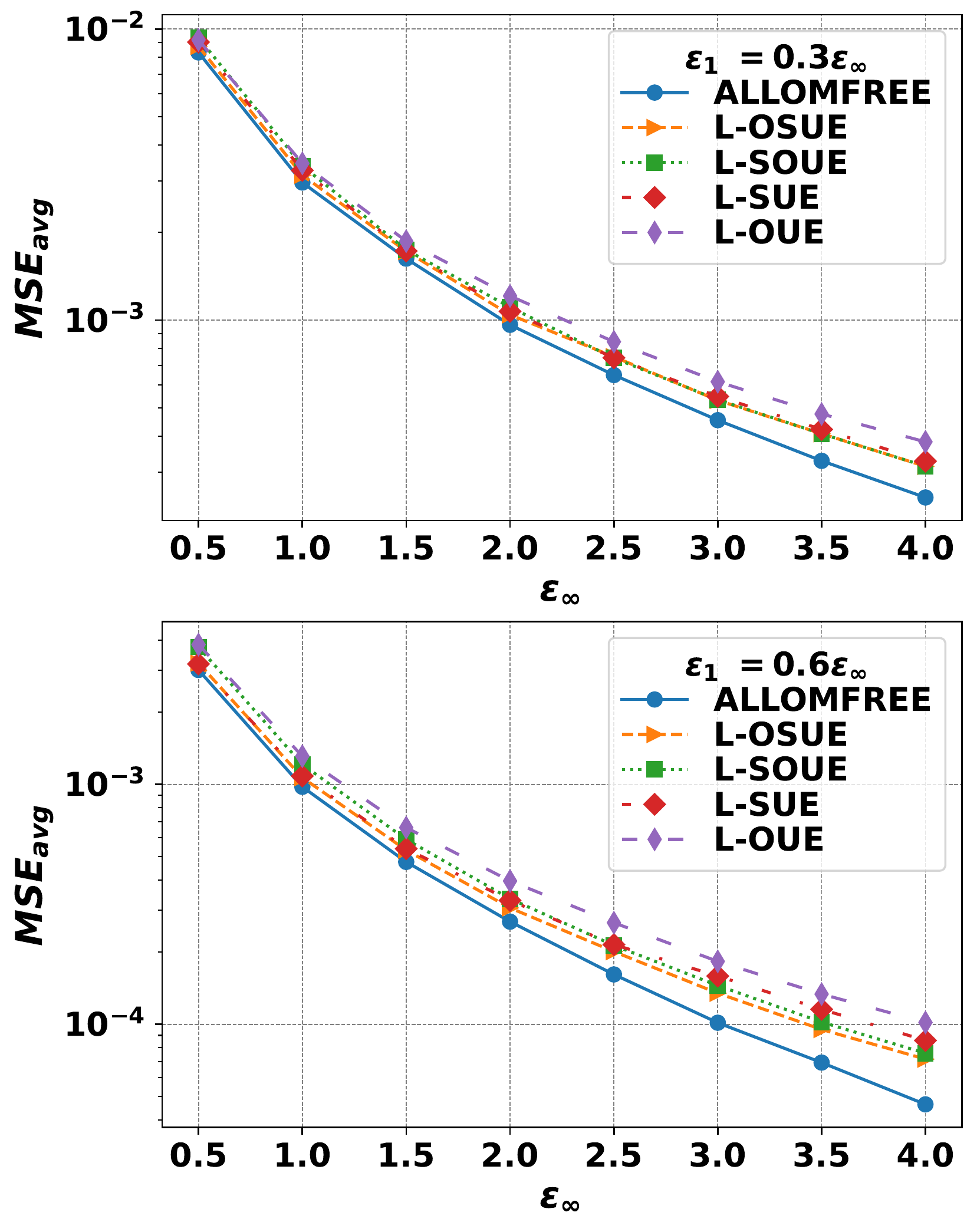}
    \caption{Averaged MSE varying $\epsilon_{\infty}$ with $\epsilon_1 = 0.3\epsilon_{\infty}$ (plot on the top) and with $\epsilon_1 = 0.6\epsilon_{\infty}$ (plot on the bottom) on the \textit{Census-Income} dataset.}% with $n=299285$, $d=33$, and domain size \begin{math}\textbf{k}=[9, 52, 47, 17,  3,  ..., 43, 43, 43,  5,  3,  3,  3,  2]\end{math}
    \label{fig:results_census}
\end{figure}

Our experiments were conducted on four real-world datasets with varied parameters for $n$, $d$, and $\textbf{k}$, which allowed evaluating our solutions more practically. Fig.~\ref{fig:results_nursery} (\textit{Nursery}), Fig.~\ref{fig:results_adults} (\textit{Adult}), Fig.~\ref{fig:results_vhs} (\textit{MS-FIMU}), and Fig.~\ref{fig:results_census} (\textit{Census-Income}) illustrate for all the evaluated protocols, the averaged $MSE_{avg}$ (y-axis) according to the longitudinal privacy parameter $\epsilon_{\infty}$ (x-axis) with $\epsilon_1 = 0.3\epsilon_{\infty}$ (plot on the top) and with $\epsilon_1 = 0.6\epsilon_{\infty}$ (plot on the bottom), respectively.

As one can notice in the results, for all datasets, ALLOMFREE consistently and considerably outperforms the state-of-the-art protocols, namely, L-SUE (a.k.a. Basic-RAPPOR)~\cite{rappor} and L-OUE (that uses OUE~\cite{tianhao2017} twice). Indeed, the difference between the performances of ALLOMFREE and the other longitudinal LDP protocols increases proportionally according to the privacy guarantees, i.e., for high $\epsilon_{\infty}$ and $\epsilon_1$ values, the gap is bigger. This is first because in all datasets there are attribute(s) with a small domain size (e.g., $k_j=2$ or $k_j=3$), in which L-GRR can provide smaller variance values than the L-UE protocols (cf. subsection~\ref{sub:analysis_long}). Secondly, by adequately selecting the probabilities $p_1,q_1,p_2,q_2$ for the L-UE protocol (i.e., L-OSUE) also optimizes data utility. Thus, since there is a way to measure the approximate variance of the extended protocols (i.e., Eq.~\eqref{var:aprox_longitudinal}), given the sampled attribute, ALLOMFREE adaptively selects one of the optimized protocol (i.e., L-GRR or L-OSUE) whose smaller variance improves the data utility.

In addition, among the L-UE protocols applied individually, the experimental results with multidimensional data approximate the numerical results with a single attribute from subsection~\ref{sub:analysis_long}. For instance, the proposed L-OSUE provides similar or better performance than L-SUE while always outperforming L-OUE. Besides, L-SOUE always outperforms L-OUE too, achieving performance similar to those of L-OSUE and L-SUE in low privacy regimes (i.e., high $\epsilon$ values). As we have already shown in subsection~\ref{sub:analysis_long}, even though OUE has better utility than SUE for one-time collection~\cite{tianhao2017}, applying OUE twice does not provide higher utility.

To complement the results of Figs.~\ref{fig:results_nursery} --~\ref{fig:results_census}, Table~\ref{tab:utility_results_03} ($\epsilon_1 = 0.3\epsilon_{\infty}$) and Table~\ref{tab:utility_results_06} ($\epsilon_1 = 0.6\epsilon_{\infty}$) exhibit all datasets and $\epsilon_{\infty}$ guarantees the following utility metrics:

\begin{equation} \label{eq:utility}
\begin{gathered}
    \mathcal{U}_{\textit{L-SUE}} = \frac{MSE_{avg_{(\textit{L-SUE})}} - MSE_{avg_{(\textit{ALLOMFREE})}}}{MSE_{avg_{(\textit{L-SUE})}}}  \\
    \mathcal{U}_{\textit{L-OUE}} = \frac{MSE_{avg_{(\textit{L-OUE})}} - MSE_{avg_{(\textit{ALLOMFREE})}}}{MSE_{avg_{(\textit{L-OUE})}}} 
\end{gathered}
\end{equation}

\noindent in which $\mathcal{U}_{\textit{L-SUE}}$ and $\mathcal{U}_{\textit{L-OUE}}$ represent the accuracy gain of ALLOMFREE over the state-of-the-art L-SUE and L-OUE protocols, respectively.

\setlength{\tabcolsep}{2pt}
\begin{table}[t]
    \scriptsize
    \centering
    \begin{tabular}{lrrrrrrrr}
    \toprule
    \multirow{2}{*}{$\epsilon_{\infty}$} & \multicolumn{2}{c}{\textit{Nursery}}  &  \multicolumn{2}{c}{\textit{Adult}}  &  \multicolumn{2}{c}{\textit{MS-FIMU}}  &  \multicolumn{2}{c}{\textit{Census-Income}}  \\ 
    & $\mathcal{U}_{\textit{L-SUE}}$ & $\mathcal{U}_{\textit{L-OUE}}$ & $\mathcal{U}_{\textit{L-SUE}}$ & $\mathcal{U}_{\textit{L-OUE}}$ & $\mathcal{U}_{\textit{L-SUE}}$ & $\mathcal{U}_{\textit{L-OUE}}$ & $\mathcal{U}_{\textit{L-SUE}}$ & $\mathcal{U}_{\textit{L-OUE}}$ \\
    \midrule
    0.5 &  13.51 &  20.63 &  19.03 &  27.73 &   3.03 &   5.43 &   7.84 &   9.48 \\
    1.0 &  12.36 &  17.75 &  12.77 &  20.44 &   1.01 &  11.57 &   9.21 &  14.08 \\
    1.5 &  19.95 &  25.86 &   8.47 &  18.01 &   4.13 &  11.55 &   5.82 &  12.92 \\
    2.0 &  17.18 &  33.24 &   4.11 &  17.16 &  13.22 &  23.44 &  10.06 &  20.41 \\
    2.5 &  20.70 &  35.40 &  11.93 &  22.54 &  10.41 &  22.25 &  12.77 &  23.15 \\
    3.0 &  28.69 &  42.98 &   8.35 &  28.22 &  13.07 &  21.56 &  17.07 &  26.21 \\
    3.5 &  36.19 &  54.02 &  18.97 &  32.02 &  14.78 &  29.10 &  22.02 &  30.96 \\
    4.0 &  41.24 &  57.16 &  19.81 &  34.25 &  20.38 &  29.64 &  24.99 &  35.60 \\\hline
    Mean &  23.73 &  35.88 &  12.93 &  25.05 &  10.00 &  19.32 &  13.72 &  21.60 \\
    \bottomrule
    \end{tabular}
    \caption{Accuracy gain of ALLOMFREE over the state-of-the-art L-SUE and L-OUE protocols for all datasets with $\epsilon_1 = 0.3\epsilon_{\infty}$, measured with the $\mathcal{U}_{\textit{L-SUE}}$ and $\mathcal{U}_{\textit{L-OUE}}$ metrics expressed in $\%$.}
    \label{tab:utility_results_03}
\end{table}

\setlength{\tabcolsep}{2pt}
\begin{table}[t]
    \scriptsize
    \centering
    \begin{tabular}{lrrrrrrrr}
    \toprule
    \multirow{2}{*}{$\epsilon_{\infty}$} & \multicolumn{2}{c}{\textit{Nursery}}  &  \multicolumn{2}{c}{\textit{Adult}}  &  \multicolumn{2}{c}{\textit{MS-FIMU}}  &  \multicolumn{2}{c}{\textit{Census-Income}}  \\ 
    & $\mathcal{U}_{\textit{L-SUE}}$ & $\mathcal{U}_{\textit{L-OUE}}$ & $\mathcal{U}_{\textit{L-SUE}}$ & $\mathcal{U}_{\textit{L-OUE}}$ & $\mathcal{U}_{\textit{L-SUE}}$ & $\mathcal{U}_{\textit{L-OUE}}$ & $\mathcal{U}_{\textit{L-SUE}}$ & $\mathcal{U}_{\textit{L-OUE}}$ \\
    \midrule
    0.5 &  17.82 &  38.84 &  10.42 &  27.46 &   6.41 &  24.79 &   5.65 &  21.61 \\
    1.0 &  14.99 &  38.97 &   9.83 &  25.14 &   2.97 &  23.32 &   9.79 &  25.46 \\
    1.5 &  15.88 &  41.05 &  12.90 &  28.59 &  16.00 &  30.52 &  11.88 &  28.05 \\
    2.0 &  27.52 &  54.69 &  12.95 &  33.78 &  14.81 &  35.65 &  18.45 &  32.31 \\
    2.5 &  39.59 &  60.96 &  23.28 &  38.50 &  17.71 &  35.34 &  24.89 &  39.11 \\
    3.0 &  40.64 &  65.32 &  28.59 &  47.95 &  27.26 &  40.97 &  36.12 &  44.48 \\
    3.5 &  44.39 &  68.73 &  34.85 &  50.00 &  33.69 &  50.94 &  40.01 &  48.18 \\
    4.0 &  42.24 &  71.13 &  45.26 &  58.33 &  41.83 &  59.47 &  45.85 &  54.44 \\\hline
    Mean &  30.38 &  54.96 &  22.26 &  38.72 &  20.08 &  37.62 &  24.08 &  36.70 \\
    \bottomrule
    \end{tabular}
    \caption{Accuracy gain of ALLOMFREE over the state-of-the-art L-SUE and L-OUE protocols for all datasets with $\epsilon_1 = 0.6\epsilon_{\infty}$, measured with the $\mathcal{U}_{\textit{L-SUE}}$ and $\mathcal{U}_{\textit{L-OUE}}$ metrics expressed in $\%$.}
    \label{tab:utility_results_06}
\end{table}

From Tables~\ref{tab:utility_results_03} and~\ref{tab:utility_results_06}, one can notice that ALLOMFREE considerably improves the quality of the frequency estimates in comparison with the state-of-the-art L-SUE and L-OUE protocols. On average, ALLOMFREE improves the results of L-SUE at least $10\%$ with the \textit{MS-FIMU} dataset in Table~\ref{tab:utility_results_03} and at most $30.38\%$ with the \textit{Nursery} dataset in Table~\ref{tab:utility_results_06} for the privacy guarantees $\epsilon_{\infty}$ and $\epsilon_1$ analyzed. Similarly, on average, ALLOMFREE improves the results of L-OUE at least $19.32\%$ with the \textit{MS-FIMU} dataset in Table~\ref{tab:utility_results_03} and at most $54.96\%$ with the \textit{Nursery} dataset in Table~\ref{tab:utility_results_06}. The highest gain of accuracy was about $\sim 71\%$, achieved with the \textit{Nursery} dataset when $\epsilon_{\infty}=4$ in Table~\ref{tab:utility_results_06} in comparison with the L-OUE protocol. Finally, as one can note, with higher values of $\epsilon_1$, ALLOMFREE will provide much higher utility than the other protocols.

\section{Related work} \label{sec:rel_work}

In recent times, there have been several studies on the local DP setting in both academia~\cite{Murakami2019,wang2019,xiao2,first_ldp,tianhao2017,Alvim2018,Hadamard,Bassily2015,Xiong2020,kairouz2016discrete,Wang2021_b,Naor2020,Cormode2021,Arcolezi_rs_fd} and practical deployment~\cite{rappor,microsoft,apple,Kessler2019}. The local DP model does not rely on collecting raw data anymore, which has a clear connection with the concept of randomized response~\cite{Warner1965}. Among many other complex tasks (e.g., heavy hitter estimation~\cite{Bassily2015,Wang2021,bassily2017practical}, machine learning~\cite{Chamikara2020,zhou2021local}, frequent itemset mining~\cite{Wang2018,Qin2016}), frequency estimation is a fundamental primitive in LDP and has received considerable attention for a single attribute~\cite{Cormode2021,Murakami2019,Hadamard,Wang2021_b,tianhao2017,kairouz2016discrete,Alvim2018,rappor,microsoft,Kim2018,Arcolezi2020,Zhao2019,Li2020,Wang2017}.

However, most studies for collecting multidimensional data with LDP mainly focused on numerical data~\cite{Xiong2020} (e.g.,~\cite{xiao2,wang2019,Duchi2018,Wang2021_b}) or other complex tasks with categorical data (e.g., marginal estimation~\cite{Shen2021,Peng2019,Zhang2018,Ren2018,Fanti2016}, analytical/range queries~\cite{Jianyu2020,Xu2020,Gu2019,Cormode2019}). Our ALLOMFREE solution is based on the multidimensional \textit{Smp} solution, which randomly samples a single attribute per user only, minimizing the variance of the estimation and the communication cost. A recent study~\cite{Arcolezi_rs_fd} proposes the Random Sampling plus Fake Data (RS+FD) solution for multidimensional data, in which the user samples a single attribute, but also generates fake data for all non-sampled attributes. The RS+FD solution creates uncertainty in the view of the aggregator while achieving similar data utility as the \textit{Smp} solution. An interesting direction would be to extend ALLOMFREE to add fake data for non-sampled attributes too.

Besides, most academic literature on frequency estimation focuses on single data collection. To address longitudinal data collections, in~\cite{rappor,microsoft}, the authors proposed LDP protocols based on two rounds of sanitization, i.e., \textit{memoization}, which was also adopted in this paper. In the literature, some studies~\cite{Kim2018,Vidal2020} applied L-SUE (a.k.a. Basic-RAPPOR~\cite{rappor}) and L-OUE (i.e., OUE~\cite{tianhao2017} with memoization) for longitudinal frequency estimates. However, rather than strictly using only SUE or OUE, we prove that the optimal combination is to start with OUE and then with SUE (i.e., L-OSUE). The privacy guarantees of chaining two LDP protocols has been further studied in~\cite{Naor2020,erlingsson2020encode}, which results in Eq.~\eqref{eq:eps_long}. Indeed, combining ``multiple" settings (i.e., many attributes and several collections throughout time) imposes several challenges, for which this paper proposes the first solution named ALLOMFREE under LDP.

\section{Conclusion}\label{sec:conc}

This paper investigates the problem of collecting multidimensional data throughout time for the fundamental task of frequency estimation under LDP guarantees. We extend and analyze three state-of-the-art LDP protocols, namely, GRR~\cite{kairouz2016discrete}, OUE~\cite{tianhao2017}, and SUE~\cite{rappor}, and propose an optimized solution, namely, ALLOMFREE, which randomly samples one attribute per user and adaptively selects a protocol with a lower variance (i.e., L-GRR or L-OSUE) in order to improve data utility. Through experimental validations, we demonstrate the advantages of ALLOMFREE over the state-of-the-art protocols L-SUE~\cite{rappor} and L-OUE~\cite{tianhao2017} by using four real-world datasets, with the gain of accuracy on average ranging from $10\%$ up to $55\%$ for the analyzed range of $\epsilon_{\infty}$ and $\epsilon_1$ privacy guarantees. For future work, we suggest and intend to improve the frequency estimates through post-processing techniques~\cite{ElSalamouny2020,Wang2020_post_process} and to design LDP protocols for longitudinal and multidimensional studies considering both numerical and categorical data.

\section*{Acknowledgements}
\noindent This work was supported by the EIPHI-BFC Graduate School (contract ``ANR-17-EURE-0002") and by the Region of Bourgogne Franche-Comt\'e CADRAN Project. The work of H\'eber H. Arcolezi was partially supported by the European Research Council (ERC) project HYPATIA under the European Union’s Horizon 2020 research and innovation programme. Grant agreement n. 835294. All computations have been performed on the \enquote{M\'esocentre de Calcul de Franche-Comt\'e}.

\bibliographystyle{elsarticle-num}
\bibliography{ms.bib}
~~~\\
~~~\\

\end{document}